\documentclass[prd,aps,showpacs,nofootinbib,amsmath,amssymb]{revtex4}
\usepackage{amssymb,epsfig}
\usepackage[usenames,dvipsnames]{color}
\usepackage[bookmarks]{hyperref}
\usepackage[usenames,dvipsnames]{color}

\newcommand{\Real}{\mathbb{R}}
\newcommand{\Complex}{\mathbb{C}}
\newcommand{\re}{\mbox{Re}}
\newcommand{\im}{\mbox{Im}}
\newcommand{\const}{\text{const}}

\begin{document}

\title{Quasi-normal acoustic oscillations in the Michel flow}

\author{Eliana Chaverra$^{1,2}$, Manuel D. Morales$^{1,3}$ and Olivier Sarbach$^{1,2,3}$}
\affiliation{$^1$Instituto de F\'isica y Matem\'aticas,
Universidad Michoacana de San Nicol\'as de Hidalgo,
Edificio C-3, Ciudad Universitaria, 58040 Morelia, Michoac\'an, M\'exico,\\
$^2$Gravitational Physics, Faculty of Physics, University of Vienna, Boltzmanngasse 5, 1090 Vienna, Austria,\\
$^3$Perimeter Institute for Theoretical Physics, 31 Caroline Street, Waterloo, Ontario N2L 2Y5, Canada.}

\begin{abstract}
We study spherical and nonspherical linear acoustic perturbations of the Michel flow, which describes the steady radial accretion of a perfect fluid into a nonrotating black hole. The dynamics of such perturbations are governed by a scalar wave equation on an effective curved background geometry determined by the acoustic metric, which is constructed from the spacetime metric and the particle density and four-velocity of the fluid. For the problem under consideration in this article the acoustic metric has the same qualitative features as an asymptotically flat, static and spherically symmetric black hole, and thus it represents a natural astrophysical analogue black hole.

As for the case of a scalar field propagating on a Schwarzschild background, we show that acoustic perturbations of the Michel flow exhibit quasi-normal oscillations. Based on a new numerical method for determining the solutions of the radial mode equation, we compute the associated frequencies and analyze their dependency on the mass of the black hole, the radius of the sonic horizon and the angular momentum number. Our results for the fundamental frequencies are compared to those obtained from an independent numerical Cauchy evolution, finding good agreement between the two approaches. When the radius of the sonic horizon is large compared to the event horizon radius, we find that the quasi-normal frequencies scale approximately like the surface gravity associated with the sonic horizon.
\end{abstract}

\date{\today}


\maketitle

\section{Introduction}

The study of accretion into a black hole plays a very important role in general relativity and astrophysics. In particular, an understanding of the emission of electromagnetic radiation generated by compression or friction in the gas is an important subject since this radiation may carry information about the spacetime geometry close to the black hole and thus offer the opportunity to test Einstein's general theory of gravity in its strong field limit. In fact, millimeter-wave very-long baseline interferometric arrays such as the Event Horizon Telescope~\cite{EHT} are already able to resolve the region around Sagittarius A$^*$, the supermassive black hole lying in the center of our galaxy, to scales smaller than its gravitational radius~\cite{sDetal08}. Comparing the observations to calculated images of the black hole shadow and the sharp photon ring surrounding it may even lead to tests for the validity of the no-hair theorems~\cite{aBtJaLdP14}.

Clearly, the features of the observed electromagnetic signals depend on the properties and dynamics of the flow, and therefore it is of considerable interest to study the dynamics of the accreted gas and to identify its key properties like its oscillation modes, for example. For a numerical study of oscillating relativistic fluid tori around a Kerr black hole and astrophysical implications, see Ref.~\cite{oZjFlRpM05}. For the impact of a binary black hole merger on the dynamics of the circumbinary disk and associated electromagnetic signals, see Refs.~\cite{mMetal09,mAetal10} and references therein.

Motivated by the above considerations, the purpose of the present work is to study the oscillation modes of a simple accretion model, namely the radial flow of a perfect fluid on a nonrotating black hole background. Spherically symmetric steady-state configurations in this model for which the density is nonzero and the matter is at rest at infinity have been studied long time ago by Michel~\cite{fM72}, generalizing previous work by Bondi~\cite{hB52} in the Newtonian case. The Michel flow describes a transonic flow, the flow's radial velocity measured by static observers being subsonic in the asymptotic region and supersonic close to the event horizon. Although much less realistic than the case where the black hole rotates and/or the matter has an intrinsic angular momentum, resulting in an accretion disk, the study of spherical accretion is still relevant in a variety of interesting astrophysical scenarios. Examples include nonrotating black holes accreting matter from the interstellar medium~\cite{fM72,Shapiro-Book} and supermassive black holes accreting dark matter~\cite{fGfL11}. For a rigorous treatment on the Michel flow and its generalization to a wide class of spherical black hole backgrounds, we refer the reader to our recent work~\cite{eCoS15a}.

In this article, we study spherical and nonspherical linear acoustic perturbations of the Michel flow, assuming a fixed Schwarzschild black hole background. Moncrief~\cite{vM80} showed that if the entropy and vorticity perturbations are of bounded extent on some initial hypersurface, they will be advected into the black hole in finite time, leaving a pure potential flow perturbation in their wake. Furthermore, Moncrief showed in Ref.~\cite{vM80} that the potential flow perturbation can be described in a very elegant manner by a wave equation on an effective curved background geometry described by the \emph{acoustic (or sound) metric}, which is constructed from the spacetime metric and the four-velocity and particle density of the background flow. The acoustic metric is Lorentzian and its null cones (the sound cones) lie inside the light cones, as long as the speed of sound is smaller than the speed of light. For further properties of the acoustic metric, see~\cite{nB99}.

For acoustic perturbations of the Michel flow the geometry described by the acoustic metric is asymptotically flat, static and spherically symmetric and possesses a sonic horizon, defined as the boundary of the region which can send sound signals to a distant observer, where the matter is almost at rest. As it turns out, this boundary coincides with the location of the sonic sphere describing the transition of the flow's radial velocity measured by static observers from sub- to supersonic. Therefore, as far as the propagation of sound waves are concerned, the acoustic geometry for the Michel flow has exactly the same qualitative properties as the geometry of a static, spherically symmetric black hole on which electromagnetic radiation propagates, and the sonic horizon in the acoustic geometry plays the role of the event horizon. Consequently, the acoustic geometry for the Michel flow constitutes a natural astrophysical ``analogue black hole". For a review on analogue black holes in different physical contexts, we refer the reader to Ref.~\cite{cBsLmV05}, and for recent applications to accretion flows on black hole backgrounds, see Refs.~\cite{pMeM13,tDnBsD06,tD07,dAsBtD14a,dAsBtD14b}.

Interpreting the acoustic perturbations as an evolution problem on an effective geometry leads to new insight and new results. For the case of the Michel flow, for example, one can prove that acoustic perturbations outside the sonic horizon stay bounded, using standard energy conservation techniques~\cite{vM80,pMeM13,dAsBtD14a}. In this article, we use this analogue black hole interpretation and show that, similar to the case in which a Schwarzschild black hole is perturbed, small perturbations of the Michel flow lead to \emph{quasi-normal acoustic oscillations} characterized by complex frequencies $s = \sigma + i\omega$, where $\sigma < 0$ describes the decay rate and $\omega$ the frequency of oscillation. As in the black hole case, these frequencies describe the ringdown phase which is taken over by a slower power-law decay at late times. We numerically compute the quasi-normal frequencies (and in some cases also the exponent in the late-time power-law tail) as a function of the black hole mass (or its Schwarzschild radius $r_H$), the radius of the sonic horizon $r_c$ and the angular momentum number $\ell$ of the perturbation. For previous studies of quasi-normal oscillations in fluid analogue black hole modes, see for example Refs.~\cite{eBvCjL04,cBsLmV05,sDlOlC10}. Contrary to these references which are mainly concerned with analogue black holes in the laboratory, the scenario considered in this article refers to an \emph{astrophysical} analogue black hole.

The remainder of this work is organized as follows. In Sec.~\ref{Sec:Michel} we briefly review the main features of the Michel flow, and in particular we discuss the properties of the flow in the vicinity of the sonic sphere. Next, in Sec.~\ref{Sec:QNMMode} we first analyze the geometric properties of the acoustic metric and show that it indeed describes an analogue black hole whose horizon is located at the sonic sphere. We also compute the surface gravity associated with this sonic horizon since it plays an important role in the description of the quasi-normal acoustic frequencies found in this article. Next, by performing a mode decomposition, we reduce the wave equation on the acoustic metric background to a family of radial, time-independent Schr\"odinger-like equations and discuss our method for computing the quasi-normal frequencies. One important issue we would like to point out here is that unlike the case where the background metric is Schwarzschild, the effective potential appearing in our radial equation cannot be written in explicit form. This complication stems from the fact that the Michel solution, describing the particle density as a function of the areal radius coordinate, is only known in implicit form, and consequently the metric coefficients in the acoustic metric and the effective potential in the radial equation can only be described in terms of implicit functions. For this reason the problem is much harder than in the Schwarzschild case, and popular analytic methods based on series expansions like Leaver's method~\cite{MQNdeS} do not seem feasible. This issue has motivated us to reconsider the problem of calculating the quasi-normal frequencies based on a new numerical matching procedure, where the local solutions of the radial equation which are being matched are computed via a Banach iteration method. This method, which shares some common features with the complex coordinate WKB approximation (see~\cite{nAcH04} and references therein), is described and tested in Sec.~\ref{Sec:QNMMode}. See also~\cite{aZhYmZyClL14} for a recent method allowing to compute the quasi-normal frequencies for deformed Kerr black holes based on ideas from perturbation theory in quantum mechanics.

Next, in Sec.~\ref{Sec:QNMCauchy} we describe a completely different method for computing the quasi-normal frequencies based on a numerical Cauchy evolution of the wave equation. In this method, one specifies an initial perturbation for the fluid's acoustic potential, solves the wave equation numerically and registers the signal observed by a static observer outside the sonic horizon. The signal reveals an initial burst followed by a ringdown signal whose oscillations frequency $\omega$ and decay rate $\sigma$ can be combined into a complex frequency. Comparing $s = \sigma + i\omega$ with the fundamental quasi-normal frequencies computed in Sec.~\ref{Sec:QNMMode} provides a further validation for our matching procedure, and shows that the quasi-normal acoustic oscillations found in this paper are actually excited by an initial perturbation of the fluid. The numerical results in Sec.~\ref{Sec:QNMCauchy} also indicate that the ringdown signal is overtaken by a power-law decay at late times, similar to what has been observed in laboratory-type analogue black holes~\cite{sDlOlC10}.

Our main results for the quasi-normal acoustic frequencies and their dependency on $r_c$ and $r_H$ and on the angular momentum number $\ell$ are presented in Sec.~\ref{Sec:Results} for the case of a polytropic fluid equation of state with adiabatic index $\gamma = 4/3$. Our results indicate that for large $r_c/r_H$ the complex frequencies $s$ scale approximately like the surface gravity $\kappa$ of the acoustic geometry and that the rescaled decay rates $\sigma/\kappa$ do not depend strongly on $\ell$ for $r_c\gg r_H$ and $\ell\geq 1$. Results for overtone frequencies and the eikonal limit $\ell\to\infty$ are also discussed in Sec.~\ref{Sec:Results}. Conclusions are drawn in Sec.~\ref{Sec:Conclusions} and technical details related to the analytic continuation of the effective potential needed for our matching procedure are explained in an appendix.

\section{Review of Michel flow and its relevant properties}
\label{Sec:Michel}

In this section, we review the relevant equations describing the Michel flow on a Schwarzschild background. For details and a generalization to more general static, spherically symmetric black hole backgrounds, see Refs.~\cite{eCoS12,eCoS15a,jKeM13}. We write the Schwarzschild metric in the form
\begin{equation}
{\bf g} = -N(r) c^2 dt^2  + \frac{dr^2}{N(r)}
 + r^2\left( d\vartheta^2 + \sin^2\vartheta d\varphi^2 \right),\qquad
 N(r) = 1 - \frac{r_H}{r},
\label{Eq:metric}
\end{equation}
where $c$ is the speed of light and $r_H$ the Schwarzschild radius. The fluid is described by the particle density $n$, energy density $\varepsilon$ and pressure $p$ measured by an observer moving along the fluid four-velocity ${\bf u} = u^\mu\partial_\mu$. (${\bf u}$ is normalized such that $u^\mu u_\mu = -c^2$.) Its dynamics is determined by the equations of motion
\begin{equation}
\nabla_\mu J^\mu=0,
\label{Eq:consCorri}
\end{equation}
 \begin{equation}
\nabla_\mu T^{\mu\nu}=0.
\label{Eq:consEner}
\end{equation}
wherein $J^\mu = n u^\mu$ is the particle current density and $T^{\mu\nu} = n h u^\mu u^\nu + p g^{\mu\nu}$ is the stress-energy tensor, and $\nabla$ refers to the covariant derivative with respect to the spacetime metric ${\bf g}$. Here and in the following, $h$ denotes the enthalpy per particle, defined as $h:= (p + \varepsilon)/n$, and we assume that $h = h(n)$ is a function of the particle density $n$ only. In the spherically symmetric stationary case Eqs.~(\ref{Eq:consCorri},\ref{Eq:consEner}) reduce to 
\begin{eqnarray}
4\pi r^2 n u^r &=& j_n = \const,
\label{Eq:jn}\\
4\pi r^2 n h u^r\sqrt{N + \left( \frac{u^r}{c} \right)^2} &=& j_\varepsilon = \const,
\label{Eq:je}
\end{eqnarray}
which expresses the conservation of particle and energy flux through a sphere of constant areal radius $r$. Using Eq.~(\ref{Eq:jn}) in order to eliminate $u^r$ in Eq.~(\ref{Eq:je}) gives
\begin{equation}
F(r,n) := h(n)^2 \left[ N(r) + \frac{\mu^2}{r^4 n^2} \right]
 = \left(\frac{j_\varepsilon}{j_n}\right)^2 = \const,\qquad
\mu := \frac{j_n}{4\pi c} < 0,
\label{Eq:Fundamental}
\end{equation}
where $j_n$ describes the accretion rate and is negative. Therefore, the problem of determining the accretion flow is reduced to finding an appropriate level curve of the function $F(r,n)$, which associates to each value of $r$ a unique value of the particle density $n(r)$. Once $n(r)$ is known, the radial velocity $u^r$ is obtained from Eq.~(\ref{Eq:jn}).

In previous work~\cite{eCoS15a} we proved that under the conditions on the equation of state (F1)--(F3) below there exists a unique smooth solution $n(r)$ of Eq.~(\ref{Eq:Fundamental}) which extends from the event horizon $r = r_H$ to infinity and has a given positive particle density $n_\infty > 0$ at infinity. We shall call this solution the Michel solution. Our conditions on $h(n)$, which we assume to be a smooth function $h: (0,\infty)\to (0,\infty)$, are the following:
\begin{enumerate}
\item[(F1)]  $h(n) \to e_0 > 0$ for $n\to 0$ (positive rest energy),
\item [(F2)] $0 < \left( \frac{v_s(n)}{c} \right)^2
 = \frac{\partial\log(h)}{\partial\log(n)} < 1$ for all  $n > 0$ (positive and subluminal sound velocity),
\item [(F3)] $0\leq W(n):=\frac{\partial\log v_s}{\partial\log n}\leq \frac{1}{3}$ for all $n > 0$ (technical restriction on the derivative of $v_s(n)$).
\end{enumerate}
In particular, these conditions are satisfied for a polytropic equation of state
\begin{equation}
h(n) = e_0 + K n^{\gamma-1},
\label{Eq:hn}
\end{equation}
wherein $e_0 > 0$, $K > 0$ and the adiabatic index $\gamma$ lies in the range $1 < \gamma \leq 5/3$. In this work, we focus on the particular case of an ultra-relativistic gas for which $h(n)$ has the same form as in Eq.~(\ref{Eq:hn}) with $\gamma = 4/3$. However, for the sake of generality, all the expressions below are given for an arbitrary equation of state satisfying the assumptions (F1)--(F3).

The function $n: [r_H,\infty) \toÊ\Real$ describing the Michel flow is a smooth, monotonously decreasing function which is implicitly determined by Eq.~(\ref{Eq:Fundamental}), that is
$$
F(r,n(r)) = \const = h(n_\infty)^2 > 0.
$$
By differentiating both sides with respect to $r$ one obtains
\begin{equation}
\frac{\partial F}{\partial r}(r,n(r)) + \frac{\partial F}{\partial n}(r,n(r)) n'(r) = 0,
\label{Eq:nprime}
\end{equation}
where the partial derivatives of $F$ are
\begin{eqnarray}
\frac{\partial F}{\partial r}(r,n) &=& \frac{h(n)^2}{r}\left[ \frac{r_H}{r} 
 - \frac{4\mu^2}{r^4 n^2} \right],
\label{Eq:Fr}\\
\frac{\partial F}{\partial n}(r,n) &=& \frac{2h(n)^2}{n}\left[ \frac{v_s^2}{c^2} N(r) 
 - \left( 1 - \frac{v_s^2}{c^2} \right)\frac{\mu^2}{r^4 n^2} \right].
\label{Eq:Fn}
\end{eqnarray}
The implicit function theorem guarantees local existence and uniqueness of $n(r)$ as long as $\partial F/\partial n \neq 0$. In the asymptotic region (large $r$), $\partial F/\partial n > 0$ is positive, and close to the event horizon ($r\simeq r_H$) $\partial F/\partial n < 0$ is negative, so in these regions the slope $n'$ of $n$ is uniquely determined by Eq.~(\ref{Eq:nprime}). However, by continuity, there exists a point $r_c > r_H$ where $\partial F/\partial n$ vanishes, and at this point $n'(r_c)$ can only be finite if $\partial F/\partial r$ also vanishes. This leads to the requirement that the flow must necessarily pass through a critical point $(r_c,n_c)$ of the function $F(r,n)$. In~\cite{eCoS15a} we proved that under the assumptions (F1),(F2),(F3) on the fluid there is, for large enough $|\mu|$, a unique critical point of $F(r,n)$ and a unique solution $n(r)$ of Eq.~(\ref{Eq:Fundamental}) which extends from $r_H$ to $r = \infty$ and satisfies $n(r_c) = n_c$. Furthermore, given $n_\infty > 0$ the value of $|\mu|$ (and hence the location of the critical point) is fixed.

Physically, the critical point corresponds to the sonic sphere $r = r_c$, which describes the transition of the flow's radial velocity measured by static observers from sub- to supersonic. The location of the sonic sphere is determined by the equations
\begin{equation}
\frac{r_c}{r_H} = \frac{1}{4}\left( 3 + \frac{1}{\nu_c^2} \right),\qquad
n_c = \frac{2|\mu|}{\sqrt{r_c^3 r_H}},\quad
\nu_c := \frac{v_s}{c},
\label{Eq:Sonic}
\end{equation}
which follow from setting the right-hand sides of Eqs.~(\ref{Eq:Fr},\ref{Eq:Fn}) to zero. According to assumption~(F2), $\nu_c^{-2}$ is always larger than one, and Eq.~(\ref{Eq:Sonic}) implies that the sonic horizon is located outside the horizon.

For later use we shall also need the derivative of the particle density $n_c' := n'(r_c)$ at the critical point. For this, we differentiate Eq.~(\ref{Eq:nprime}) with respect to $r$ and evaluate at $r = r_c$, obtaining
\begin{equation}
\frac{\partial^2 F}{\partial r^2}(r_c,n_c) + 2\frac{\partial^2 F}{\partial r\partial n}(r_c,n_c) n_c' 
 + \frac{\partial^2 F}{\partial n^2}(r_c,n_c) (n_c')^2 = 0.
\label{Eq:nprimeprime}
\end{equation}
Using the following expression for the Hessian of $F$ at $(r_c,n_c)$,
$$
\left( \begin{array}{cc}
\frac{\partial^2 F}{\partial r^2}(r_c,n_c) & \frac{\partial^2 F}{\partial r\partial n}(r_c,n_c) \\
\frac{\partial^2 F}{\partial n\partial r}(r_c,n_c) & \frac{\partial^2 F}{\partial n^2}(r_c,n_c)
\end{array} \right) =
\frac{h_c^2}{n_c^2}\frac{r_H}{r_c}\left(\begin{array}{cc}
3\frac{n_c^2}{r_c^2} & 2\frac{n_c}{r_c} \\
2\frac{n_c}{r_c} & 1 - \nu_c^2 + W_c
\end{array} \right),
$$
with $h_c = h(r_c)$ and $W_c = W(r_c)$, we find the two solutions
\begin{equation}
\frac{n_c'}{n_c} = -\frac{3}{r_c}\frac{1}{2 \pm \sqrt{1 + 3(\nu_c^2 - W_c)}},
\label{Eq:nc_prime}
\end{equation}
which parametrize the two branches of the level set of $F$ through $(r_c,n_c)$. In~\cite{eCoS15a} we proved that the branch corresponding to the global solution for $n(r)$ extending from the horizon to infinity is the one with the $+$ sign in Eq.~(\ref{Eq:nc_prime}). In the appendix, we show that the function $n(r)$ admits an analytic continuation to complex $r$. This continuation is required for the quasi-normal mode calculation in the next section.

\section{Quasi-normal oscillations from a mode analysis}
\label{Sec:QNMMode}

The propagation of acoustic perturbations in any relativistic perfect fluid is elegantly described by a wave equation
\begin{equation}
\Box_\mathfrak{G}\Psi = 0,
\label{Eq:Acoustic}
\end{equation}
where the scalar field $\Psi$ determines the perturbed enthalpy $\delta h$ and four-velocity $\delta u_\mu$ of the fluid according to the relation $\delta(h u_\mu) = \nabla_\mu\Psi$, which using $u^\mu\delta u_\mu = 0$ yields
$$
\delta h = -u^\mu\nabla_\mu\Psi,\qquad
\delta u_\mu = \frac{1}{h}\left[ \nabla_\mu\Psi + u_\mu u^\nu\nabla_\nu\Psi \right].
$$
The operator $\Box_\mathfrak{G}$ in Eq.~(\ref{Eq:Acoustic}) is the wave operator belonging to the {\em acoustic metric} $\mathfrak{G}$, which is constructed from the spacetime metric ${\bf g}$ and the fluid quantities in the following way~\cite{vM80}:
\begin{equation}
\mathfrak{G}_{\mu\nu} := \frac{n}{h}\frac{c}{v_s}\left[ g_{\mu\nu} 
 + \left(1 - \frac{v_s^2}{c^2} \right) u_\mu u_\nu \right].
\label{Eq:SoundMetric}
\end{equation}
Under our assumptions on the sound speed it follows that $\mathfrak{G}$ is a Lorentzian metric whose cone (the sound cone) lies {\em inside} the light cone of ${\bf g}$. Notice also that ${\bf u}$ is timelike with respect to both ${\bf g}$ and $\mathfrak{G}$.

\subsection{Geometry of the acoustic metric}

For simplicity, from now on we use units in which the speed of light is one, $c=1$. For the particular case of the Michel flow on a Schwarzschild metric the acoustic metric is
\begin{equation}
\mathfrak{G} = \frac{n}{h}\frac{1}{v_s}\left[ -N dt^2+ \frac{dr^2}{N} 
 + (1- v_s^2)\left(u_t dt + u_r dr \right)^2 
 + r^2\left( d\vartheta^2 + \sin^2\vartheta d\varphi^2 \right) \right],
\label{Eq:SoundMetricMichel}
\end{equation}
or
\begin{equation}
\mathfrak{G} = -A(r) dt^2 + 2B(r) dt dr + C(r) dr^2 
 + R(r)^2\left( d\vartheta^2 + \sin^2\vartheta d\varphi^2 \right),
\label{Eq:SoundMetricMichelBis}
\end{equation}
with
\begin{eqnarray*}
A(r) &=& \frac{n}{h}\frac{1}{v_s}\left[ v_s^2 N - (1 - v_s^2)(u^r)^2 \right],\\
B(r) &=& -\frac{n}{h}\frac{1-v_s^2}{v_s}\frac{\sqrt{N + (u^r)^2} u^r}{N},\\
C(r) &=& \frac{n}{h}\frac{1}{v_s}\frac{1}{N^2}\left[ N + (1-v_s^2)(u^r)^2 \right],\\
R(r) &=& \sqrt{\frac{n}{h}\frac{1}{v_s}}\, r,
\end{eqnarray*}
where we have used the equation $u_t^2 - (u^r)^2 = N$ and $u_t < 0$ in order to eliminate $u_t$ and where the quantities $n$, $h$, $v_s$ and $u^r$ are given by the Michel flow solution discussed in the previous section. The acoustic metric~(\ref{Eq:SoundMetricMichel}) is spherically symmetric and possesses the Killing vector field
\begin{equation}
k = \frac{\partial}{\partial t}
\label{Eq:KVF}
\end{equation}
whose negative square norm is $A(r)$. Since $A(r)$ is positive for $r > r_c$ and negative for $r < r_c$ (cf. Eq.~(\ref{Eq:Fn}) and the remarks following this equation) the vector field $k$ is timelike for $r > r_c$, spacelike for $0 < r < r_c$ and null at $r = r_c$, and the surface $r = r_c$ is a Killing horizon~\cite{Heusler-Book,Wald-Book}. Notice that the coordinates $(t,r)$ are regular everywhere outside the event horizon $r > r_H$; in particular they are regular at the sonic horizon $r = r_c$. Introducing the new time coordinate
$$
T := t - \int\frac{B(r)}{A(r)} dr,
$$
the acoustic metric can be brought into diagonal form outside the sonic horizon,
\begin{equation}
\mathfrak{G} = \frac{n}{h}\frac{1}{v_s}\left[ -X(r) v_s^2 dT^2 + \frac{dr^2}{X(r)}
 + r^2\left( d\vartheta^2 + \sin^2\vartheta d\varphi^2 \right) \right],\qquad
X(r) := N(r) - \left( \frac{1}{v_s^2} - 1 \right)(u^r)^2.
\label{Eq:SoundMetricMichelDiag}
\end{equation}
Note that $X(r)\to 1$ as $r\to \infty$, and in this limit the acoustic metric reduces (up to a constant conformal factor) to the Minkowksi metric with time coordinate $v_\infty T$, where $v_\infty := \lim_{r\to \infty} v_s(r)$ is the sound speed at infinity.

It follows that the geometry described by the acoustic metric~(\ref{Eq:SoundMetricMichel}) is the same as the one of a static, spherically symmetric and asymptotically flat black hole. The sonic horizon $r = r_c$ plays the role of the event horizon of this analogue black hole. Its surface gravity $\kappa$ with respect to the Killing vector field $k$ defined in Eq.~(\ref{Eq:KVF}), which will play an important role later, can be computed using Eqs.~(\ref{Eq:Sonic}) and (\ref{Eq:nc_prime}). The result is
\begin{equation}
\kappa = \frac{A'(r_c)}{2B(r_c)}
 = \frac{1}{4v_c}\frac{r_H}{r_c^2}\sqrt{1 + 3(\nu_c^2 - W_c)}.
\label{Eq:SurfaceGravity}
\end{equation}

Since scalar fields propagating on static spherically symmetric black holes like the Schwarzschild and Reissner-Nordstr\"om black holes exhibit quasi-normal oscillations, and since the fluid potential $\Psi$ satisfies a wave equation on an analogue black hole background, it is natural to expect that acoustic perturbations in the Michel flow undergo quasi-normal oscillations as well. In the following, we show that such oscillations do indeed exist and compute the associated frequencies based on two different numerical methods.

\subsection{Reduction to a Schr\"odinger-like equation}

Quasi-normal modes are particular solutions of Eq.~(\ref{Eq:Acoustic}) which are of the form
$$
\Psi = \frac{1}{R} e^{s T}\psi(s,r) Y^{\ell m}(\vartheta,\varphi),
$$
for some complex frequency $s = \sigma + i\omega\in \Complex$ and complex-valued function $\psi(s,r)$ to be determined. Here, $\sigma$ denotes the decay rate, $\omega$ the frequency of oscillations, and $Y^{\ell m}$ the standard spherical harmonics with angular momentum numbers $\ell m$. Introducing this ansatz into Eq.~(\ref{Eq:Acoustic}) and using the diagonal parametrization~(\ref{Eq:SoundMetricMichelDiag}) of the acoustic metric, one obtains the following equation:
\begin{equation}
 - {\cal N}(r)\frac{\partial}{\partial r} \left[ {\cal N}(r)\frac{\partial \psi}{\partial r} \right]
 + \left[ s^2 + {\cal N}(r) V_\ell(r)\right]\psi = 0,
\label{Eq:psi}
\end{equation}
where the functions ${\cal N}(r)$ and $V_\ell(r)$ are explicitly given by
\begin{eqnarray}
{\cal N}(r) &=& v_s X = v_s\left[ 1 - \frac{r_H}{r} - \left( \frac{1}{v_s^2} - 1 \right)(u^r)^2 \right],
\label{Eq:DefcalN}\\
V_\ell(r) &=& \frac{1}{r^2 v_s}\left\{ -(1 - v_s^2 + 5W) E
 - \frac{r}{2}\frac{n'}{n} \left[ 4W + 3W^2 + (1-v_s^2)^2 - 2\frac{dW}{d\log n} \right] E
 + \frac{r_H}{4r}\left[ 1 + 3v_s^2 + 3W + 4Wr\frac{n'}{n} \right] \right\}
\nonumber\\
 &+& v_s\frac{\ell(\ell + 1)}{r^2},
\label{Eq:DefVell}
\end{eqnarray}
with $E := r_H/(4r) - (u^r)^2$, $u^r = \mu/(r^2 n)$ and $W = \partial\log v_s/\partial\log n$. Away from the critical point $n'/n$ can be computed using Eq.~(\ref{Eq:nprime}), which yields
\begin{equation}
\frac{n'}{n} = -\frac{2}{r}\frac{E}{v_s^2 X},
\label{Eq:nprimeBis}
\end{equation}
while for $r = r_c$ Eq.~(\ref{Eq:nc_prime}) can be used in order to compute $n'/n$.

\begin{figure}[htp]
\includegraphics[width=10cm]{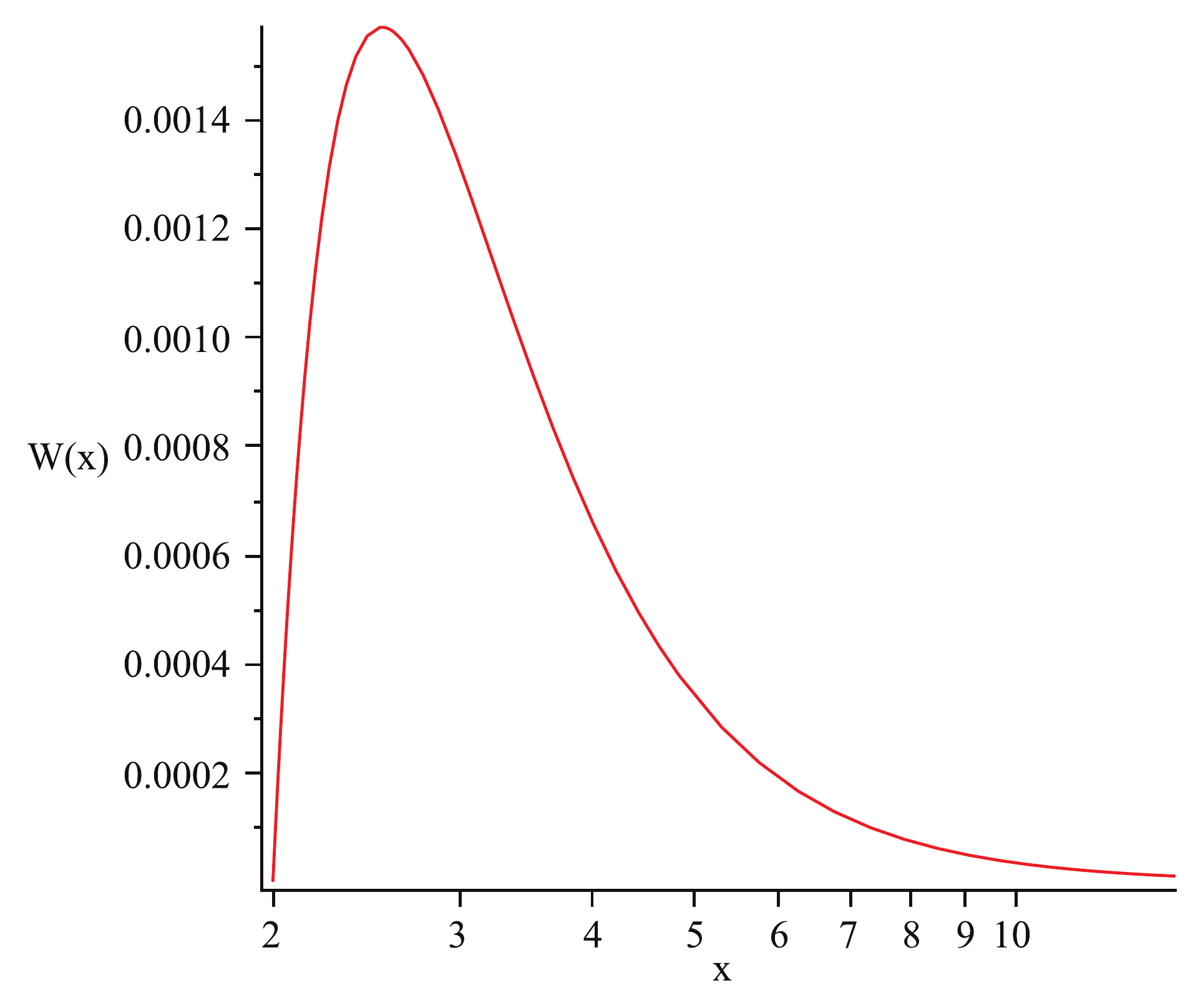}
\caption{The effective dimensionless potential $W(x) := r_H^2{\cal N}(r) V_0(r)$ in the Hamiltonian ${\cal H}$ as a function of $x := r/r_H$ is shown here for the case of the Michel flow for a polytrope with adiabatic index $\gamma = 4/3$ and sonic horizon located at $r_c = 2r_H$.}
\label{Fig:EffPot}
\end{figure}

For large $r$ the effective potential $V_\ell(r)$ behaves as $v_\infty\ell(\ell+1)/r^2 + O(r^{-3})$, so it is dominated by the centrifugal term. At the sonic horizon ${\cal N}(r_c)$ is zero, but $V_\ell(r_c)$ is positive. Introducing the tortoise coordinate $r_* = \int dr/{\cal N}(r)$ which ranges from $-\infty$ to $+\infty$ Eq.~(\ref{Eq:psi}) can be further simplified and is formally equivalent to the time-independent Schr\"odinger equation ${\cal H}\psi = -s^2\psi$ with Hamiltonian
$$
{\cal H} := -\frac{d^2}{dr_*^2} + {\cal N}(r) V_\ell(r).
$$
A plot of the effective potential ${\cal N}(r) V_\ell(r)$ for $\ell = 0$ is shown in Fig.~\ref{Fig:EffPot}, which indicates that there is a potential barrier even in the monopolar case $\ell = 0$.

\subsection{Computation of the quasi-normal frequencies using Banach iterations}

The quasi-normal frequencies $s$ are determined by the following requirement~\cite{hNbS92}. For $\sigma = \re(s) > 0$ Eq.~(\ref{Eq:psi}) admits precisely two solutions $\psi_\pm(s,r)$ satisfying the boundary conditions
\begin{equation}
\lim\limits_{r_*\to\infty} e^{s r_*}\psi_+(s,r) = 1,\qquad
\lim\limits_{r_*\to -\infty} e^{-s r_*}\psi_-(s,r) = 1,
\label{Eq:psiBC}
\end{equation}
in the asymptotic region and at the sonic horizon, respectively. These solutions can be shown to depend analytically on $s$, and they can be analytically continued on the left complex plane $\sigma < 0$. For $\sigma > 0$ the two functions $\psi_+(s,\cdot)$ and $\psi_-(s,\cdot)$ are always linearly independent from each other since otherwise one would have a finite energy solution which grows exponentially in time, in contradiction to standard energy arguments~\cite{vM80} showing the stability of the flow outside the sonic horizon. However, for particular values of the complex frequency $s = \sigma + i\omega$ with $\sigma < 0$ it is possible that the two functions become linearly dependent. These special frequencies are the ones associated with the quasi-normal modes, and as we will show in the next section they describe the ringdown phase in the dynamics of the scalar field $\Psi$. For more general discussions on quasi-normal oscillations we refer the reader to the review articles~\cite{hN99,kKbS99,eBvCaS09}.

For $\re(s) > 0$ the solutions $\psi_{\pm}$ can be constructed using the following iteration scheme,
\begin{equation}
\psi_\pm(s,r) = e^{\mp s r_*}\lim\limits_{k\to\infty} (T_{\pm s}^k 1)(r),
\label{Eq:psipm}
\end{equation}
where the operators $T_{\pm s}$, acting on continuous and bounded functions $\xi$, are defined as
\begin{eqnarray}
(T_{+s}\xi)(r) &=& 1 + \frac{1}{2s}\int\limits_r^\infty \left( 1 - e^{-2s(r_*' - r_*)} \right) 
V_\ell(r')\xi(r') dr',
\label{Eq:T+s}\\
(T_{-s}\xi)(r) &=& 1 + \frac{1}{2s}\int\limits_{r_c}^r \left( 1 - e^{+2s(r_*' - r_*)} \right) 
V_\ell(r')\xi(r') dr',
\label{Eq:T-s}
\end{eqnarray}
for $r_c < r < \infty$, with $r'$ the variable of integration and $r_*'$ the associated tortoise coordinate. Note that the integrals in these expressions are well-defined for $\sigma = \re(s)\geq 0$, because $|e^{-2s(r_*' - r_*)}| = e^{-2\sigma(r_*' - r_*)} \leq 1$ when $r_*'\geq r_*$ and because the potential $V_\ell(r)$ decays at least as fast as $1/r^2$ for $r\to\infty$. With these observations in mind it is not difficult to verify that the sequences $T_{\pm s}^k 1(r)$ obtained by applying $k$ times the operators $T_{\pm s}$ to the constant function $\xi = 1$, converge for all $\re(s)\geq 0$ with $s\neq 0$ and all $r > r_c$, uniformly on compact intervals, and that $\psi_\pm(s,\cdot)$ are solutions of Eq.~(\ref{Eq:psi}) fulfilling the required boundary conditions~(\ref{Eq:psiBC}). Furthermore, the functions $\psi_\pm(s,r)$ are analytic in $s$ for any fixed $r > r_c$. For more details on these assertions we refer the reader to Ref.~\cite{rN60} or Sec.~XI.8 in Ref.~\cite{ReedSimonIII}.

Next, let us discuss the analytic continuation of the function $\psi_+(s,r)$ for $\re(s) < 0$. In this case, the integral in Eq.~(\ref{Eq:T+s}) does not converge anymore unless $V_\ell(r)$ decays exponentially fast. However, if the effective potential $V_\ell$ and the function ${\cal N}$ in the definition of the tortoise coordinate $r_*$ possess appropriate analytic continuations on the complex $r$ plane, $\psi_+(s,r)$ can be analytically continued to $\re(s) < 0$ by deforming the path of integration in the definition of $T_{+s}\xi$ in Eq.~(\ref{Eq:T+s}). The basic idea, which has been used in Ref.~\cite{bJpC86} in the context of the Regge-Wheeler equation, relies on the following observation: for $s = |s| e^{i\varphi}$ and $r_*' - r_* = \rho e^{i\alpha}$, $\rho\geq 0$, we still have $|e^{-2s(r_*' - r_*)}| \leq 1$ as long as $\re(s(r_*' - r_*)) = |s|\rho\cos(\varphi + \alpha) \geq 0$. For the integration path in Eq.~(\ref{Eq:T+s}), $r_*'$ and $r_*$ are real and $r_*' > r_*$ and consequently, $\alpha = 0$ which implies that only those frequencies $s = \rho e^{i\varphi}$ lying in the range $|\varphi| \leq \pi/2$ (that is, $\re(s) \geq 0$) are admissible. However, choosing a new integration path such that $\alpha = -\pi/2$ leads to the admissible range $0\leq \varphi \leq \pi$, so that the integral in Eq.~(\ref{Eq:T+s}) converges for all $\im(s) > 0$ provided the analytic continuation of $V_\ell$ decays fast enough along the path (decay equal to or faster than $1/|r|^2$ is enough). Due to Cauchy's integral theorem the new integration path does yield the same value for $(T_{+s}\xi)(r)$ as the one computed using the original path in the intersection of the two domains $\re(s) > 0$ and $\im(s) > 0$, so by deforming the path in this way we obtain the required analytic continuation of $\psi_+(s,r)$ on the upper half plane $\im(s) > 0$.\footnote{A similar analytic continuation can be obtained on the lower half plane by choosing $\alpha = +\pi/2$. However, because the functions ${\cal N}$ and $V_\ell$ in Eq.~(\ref{Eq:psi}) are real the quasi-normal frequencies $s$ come in complex conjugate pairs, and thus it is sufficient to consider the upper half plane.}

In our calculations, we choose the following integration path for $T_{+s}$:
$$
\gamma_\alpha(\lambda) = r + e^{i\alpha} \lambda,\qquad \lambda\geq 0,
$$
with angle $\alpha$ slightly larger than $-\pi/2$, and set
\begin{equation}
(T_{+s}\xi)(r) = 1 + \frac{1}{2s}\int\limits_{\gamma_\alpha} \left[
  1 - \exp\left( -2s\int\limits_r^{r'} \frac{dr''}{{\cal N}(r'')} \right) \right] V_\ell(r')\xi(r') dr',\qquad
\re(r) > r_c,
\label{Eq:T+sBis}
\end{equation}
where it is understood that the integral from $r$ to $r'$ in the exponential is performed along the path $\gamma_\alpha$. The analytic continuations of the functions ${\cal N}$ and $V_\ell$ to complex $r$ and their properties are discussed in the appendix. For large $|r|$ and $\re(r) > r_c$, $V_\ell$ decays at least as fast as $1/|r|^2$ and ${\cal N}$ converges to a positive real constant, so that $r_*' - r'$ is approximately proportional to $r' - r$ for large $|r'|$. Hence the integral converges for all $\im(s) > 0$, as explained above.

The analytic continuation of the function $\psi_-(s,r)$ for $\re(s) < 0$ can be obtained using similar ideas,
\begin{equation}
(T_{-s}\xi)(r) = 1 - \frac{1}{2s}\int\limits_{\Gamma} \left[
  1 - \exp\left( 2s\int\limits_r^{r'} \frac{dr''}{{\cal N}(r'')} \right) \right] V_\ell(r')\xi(r') dr',
\label{Eq:T-sBis}
\end{equation}
with $\Gamma$ an integration path connecting $r$ with $r_c$. However, in this case particular care has to be taken regarding the relation between the tortoise coordinate and the physical radius close to the sonic horizon $r = r_c$, where the function $1/{\cal N}$ has a pole. In order to motivate our choice for the integration path $\Gamma$, we approximate
$$
\frac{1}{{\cal N}(r)} \simeq \frac{1}{{\cal N}'(r_c)}\frac{1}{r - r_c}
$$
for $r$ close to $r_c$. Note that ${\cal N}'(r_c) > 0$ is positive since the surface gravity associated with the sonic horizon is positive. As a consequence of the residual theorem, the integral over $1/{\cal N}$ increases by a factor of $2\pi i/{\cal N}'(r_c)$ after each revolution along a closed path that winds counter-clockwise around $r = r_c$. In the exponential in the integrand on the right-hand side of Eq.~(\ref{Eq:T-sBis}) this would give rise to a multiplicative factor $\exp(4\pi i s/{\cal N}'(r_c))$, which is bounded for all $\im(s) \geq 0$.

\begin{figure}[ht]
\centerline{\resizebox{6cm}{!}{\includegraphics{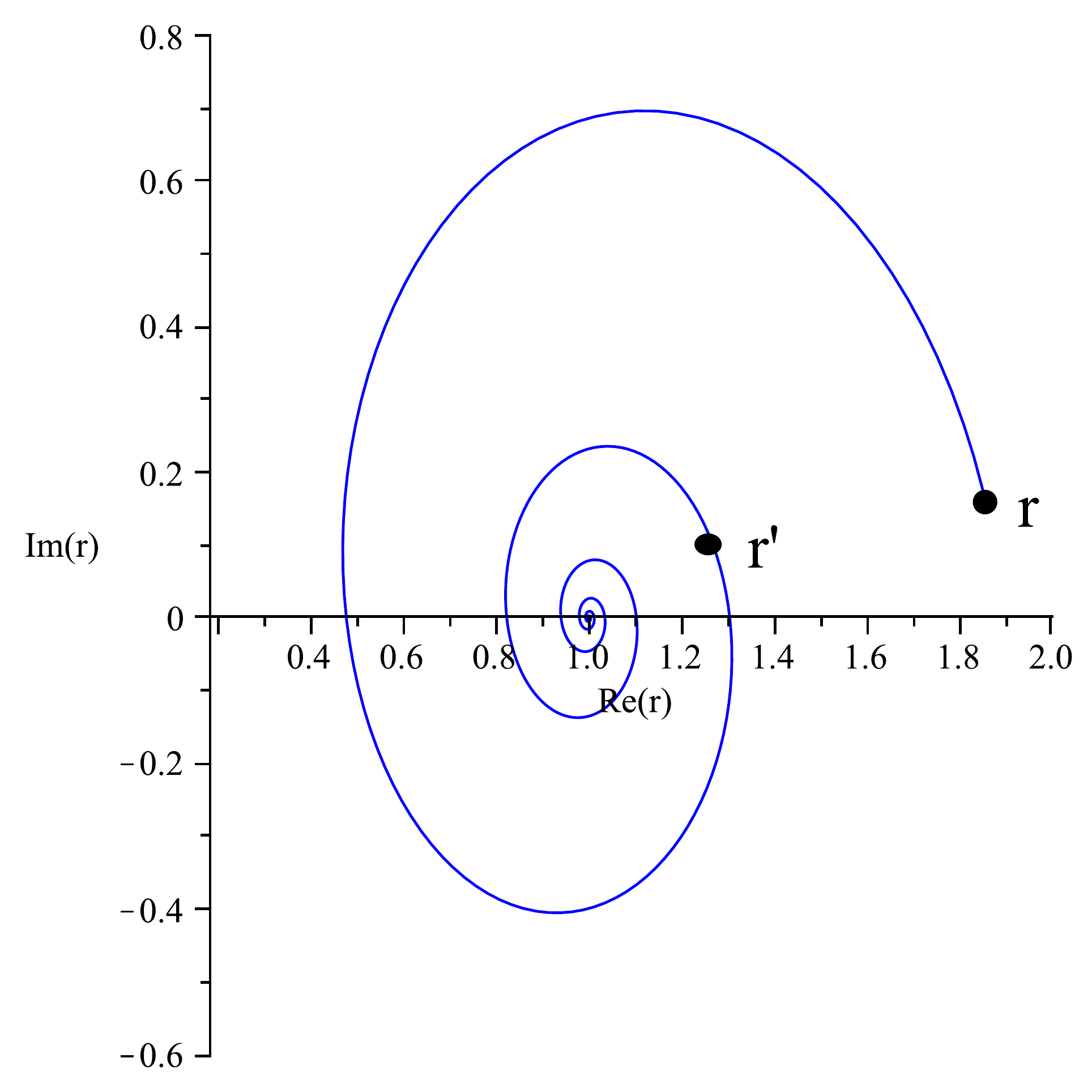}}}
\vspace{0.3cm}
\caption{The integration path $\Gamma_\beta$ in the complex $r$-plane for the case $r_c = 1$.}
\label{Fig:Spiral}
\end{figure}

Motivated by these observations, we choose the integration path
$$
\Gamma_\beta(\lambda) = r_c + (r - r_c)\exp( -e^{i\beta}\lambda ),\quad \lambda \geq 0
$$
with $\beta$ slightly larger than $-\pi/2$, which spirals counter-clockwise around the point $r = r_c$, see Fig.~\ref{Fig:Spiral}. Along this path we have, for $s = |s| e^{i\varphi}$,
$$
\exp\left( 2s\int\limits_r^{r'} \frac{dr''}{{\cal N}(r'')} \right)
 \simeq \exp\left( \frac{2s}{{\cal N}'(r_c)}\int\limits_0^{\lambda'} (-e^{i\beta}) d\lambda \right)
 = \exp\left( -2|s| e^{i(\beta + \varphi)} \frac{\lambda'}{{\cal N}'(r_c)} \right),
$$
which is bounded provided $|\beta + \varphi|\leq \pi/2$. Therefore, choosing $\beta = -\pi/2 + \delta$ with small $\delta > 0$ guarantees convergence of the integral in Eq.~(\ref{Eq:T-sBis}) for all $0 < \varphi < \pi - \delta$, so by choosing $\delta > 0$ small enough we can cover the whole upper plane $\im(s) > 0$. More details and rigorous justifications of our method will be provided elsewhere~\cite{eCoS15b}.

For given $\im(s) > 0$ we numerically compute the functions $\psi_\pm(s,r)$ and their first derivative $\psi_\pm'(s,r)$ by truncating the iteration in Eq.~(\ref{Eq:psipm}) to some finite $k$ and computing the operators $T_{\pm s}$ using Eqs.~(\ref{Eq:T+sBis}) and~(\ref{Eq:T-sBis}), where we discretize the integrals using the trapezoidal rule. We choose $\alpha = -1.57$ and $\beta = -1.5$, and we find that in practice only about $k \sim 10$ iterations are required for good accuracy. The functions ${\cal N}(r)$ and $V_\ell(r)$ are computed from Eqs.~(\ref{Eq:DefcalN},\ref{Eq:DefVell}), where $n(r)$ is determined numerically by solving Eq.~(\ref{Eq:Fundamental}) via a standard Newton algorithm~\cite{Recipes-Book}. In order to find the quasi-normal frequencies we match the two solutions $\psi_+$ and $\psi_-$ by finding the zeros of their Wronski determinant,
\begin{equation}
W(s) := \det\left( \begin{array}{rr}
\psi_+(s,r) & \psi_-(s,r)\\
{\cal N}\psi_+'(s,r) & {\cal N}\psi_-'(s,r)
\end{array} \right)
 = \xi_+(s,r){\cal N}\xi_-'(s,r) - {\cal N}\xi_+'(s,r)\xi_-(s,r) + 2s\xi_+(s,r)\xi_-(s,r),
\label{Eq:WDef}
\end{equation}
where $\xi_\pm(s,r) = \lim_{k\to\infty} (T_{\pm s}^k 1)(r)$, at some intermediate point $r_m$ ($r_c < r_m < \infty$) which we typically choose to be about $r_m \simeq 1.5r_c$. The zeros of $W$ are obtained numerically using a standard Newton algorithm~\cite{Recipes-Book}, where the derivative of $W(s)$ with respect to $s$ is approximated using a simple finite difference operator.

To test our algorithm, we have applied it to the computation of the quasi-normal frequencies for odd-parity linearized gravitational perturbations of a Schwarzschild black hole, in which case the functions ${\cal N}$ and $V_\ell$  in Eq.~(\ref{Eq:psi}) are replaced by ${\cal N}(r) = 1 - r_H/r$ and $V_\ell(r) = \ell(\ell+1)/r^2 - 3r_H/r^3$, respectively. In the quadrupolar case we found the following frequencies: $s\cdot r_H = -0.17792 + 0.74734i, -0.54783 + 0.69342i, -0.95656 + 0.60211i, -1.4103 + 0.50301i, -1.8937 + 0.41503i, -2.3912 + 0.33859i, -2.8958 + 0.26651i$, which agree to high accuracy with those obtained from Leaver's continued fraction method~\cite{MQNdeS}. In order to produce these results we have chosen $r_m = 1.5r_H$, discretized the integrals in Eqs.~(\ref{Eq:T+sBis},\ref{Eq:T-sBis}) using $40,000$ points and performed $14$ Banach iterations. We have varied these numbers in order to obtain five significant figures in all the frequencies.

\section{Quasi-normal oscillations from a Cauchy evolution}
\label{Sec:QNMCauchy}

In this section, we solve the Cauchy problem for the wave equation~(\ref{Eq:Acoustic}) numerically, starting with a Gaussian pulse with zero velocity as initial data. We show that a static observer, after registering an initial burst of radiation, measures a ringdown signal whose frequency is given by the one of the fundamental quasi-normal mode.

\subsection{Reduction to a first-order symmetric hyperbolic system}

We formulate the Cauchy problem for Eq.~(\ref{Eq:Acoustic}) on the $t = \const.$ hypersurfaces of the metric~(\ref{Eq:SoundMetricMichel}) outside the sonic horizon. To this purpose we first write the acoustic metric in its ADM form
\begin{equation}
\mathfrak{G} = -\alpha(r)^2 dt^2 + \gamma(r)^2\left( dr + \beta(r) dt \right)^2
 + R(r)^2\left( d\vartheta^2 + \sin^2\vartheta d\varphi^2 \right),
\label{Eq:SoundMetricMichelADM}
\end{equation}
with the functions $\alpha(r)$, $\beta(r)$ and $\gamma(r)$ given by
$$
\alpha(r) = \sqrt{\frac{n}{h} v_s}\frac{N}{\sqrt{Y}} ~,\qquad
\beta(r) = (1-v_s^2)\frac{N\sqrt{N + (u^r)^2}|u^r|}{Y} ~,\qquad
\gamma(r) = \sqrt{\frac{n}{h}\frac{1}{v_s}}\frac{\sqrt{Y}}{N} ~,
$$
where $Y := N + (1 - v_s^2)(u^r)^2$ and $|u^r| = |\mu|/(r^2 n)$. Using the following decomposition of $\Psi$ into spherical harmonics,
$$
\Psi = \frac{1}{r}\sum\limits_{\ell m} \phi_{\ell m}(t,r) Y^{\ell m}(\vartheta,\varphi),
$$
and introducing the auxiliary fields (suppressing the indices $\ell m$ in what follows)
$$
\pi := \frac{1}{\alpha}\left( \partial_t\phi - \beta\partial_r\phi \right) ~,\qquad
\chi := \frac{1}{\gamma}\partial_r\phi ~,
$$
Eq.~(\ref{Eq:Acoustic}) can be cast into first-order symmetric hyperbolic form:
\begin{eqnarray}
\partial_t\phi &=& \alpha\phi + \gamma\beta\chi,
\label{Eq:FOSH1}\\
\partial_t\chi &=& \frac{1}{\gamma}\partial_r( \alpha\pi + \gamma\beta\chi),
\label{Eq:FOSH2},\\
\partial_t\pi &=& \frac{1}{\gamma}\left( \frac{r}{R} \right)^2\partial_r
\left[ \left( \frac{R}{r} \right)^2(\alpha\chi + \gamma\beta\pi) \right] - \alpha U_\ell(r)\phi
\label{Eq:FOSH3},
\end{eqnarray}
with the effective potential $U_\ell(r)$ given by
\begin{equation}
U_\ell(r) = \frac{r}{\alpha\gamma R^2}\partial_r\left[ \alpha\gamma\left( \frac{R}{r} \right)^2
\left( \frac{1}{\gamma^2} - \frac{\beta^2}{\alpha^2} \right) \right]
 + \frac{\ell(\ell+1)}{R^2}.
\end{equation}
Explicit evaluation of this potential leads to
\begin{equation}
U_{\ell}(r) = \frac{2}{R^2 v_s^2}\left\{ 
 (3v_s^2 - 1) E(r) + 2(u^r)^2\left[ 1 + \frac{r}{2}\frac{n'}{n}( 1 - v_s^2 + W) \right]  \right\}
 + \frac{\ell \left(\ell +1\right)}{R^2},
\end{equation}
where we recall that $E := r_H/(4r) - (u^r)^2$ and $u^r = \mu/(r^2 n)$. As before, $n'/n$ can be computed using Eq.~(\ref{Eq:nprimeBis}) for all $r \neq r_c$ and at $r = r_c$ we can use the expression in Eq.~(\ref{Eq:nc_prime}) instead.

We solve the first-order system~(\ref{Eq:FOSH1},\ref{Eq:FOSH2},\ref{Eq:FOSH3}) using a finite-difference code based on the method of lines. The spatial domain is a finite interval $r\in [r_c,r_{out}]$ with $r_{out} \gg r_c$ large enough such that spurious reflections from the outer boundary do not affect the wave signal measured by the static observer for the times used in our simulations. There are no boundary conditions that must be specified at the inner boundary $r = r_c$ since there all the characteristic velocities
$$
\lambda_0 = 0,\qquad
\lambda_\pm = \beta \pm \frac{\alpha}{\gamma}
 = \frac{N}{Y}\left[ \pm v_s N - (1 - v_s^2)\sqrt{N + (u^r)^2} u^r \right]
$$
are zero or positive. At the outer boundary $r = r_{out}$ there is one incoming mode
$$
v_{in} = \frac{1}{\sqrt{2}}(\pi + \chi)
$$
which we set to zero. Although this boundary condition is not exactly transparent to the physical problem, it yields only small spurious reflections when $r_{out} \gg r_H$ and as mentioned above, we extract the physical information only at events which are causally disconnected from the boundary surface in order to make sure that there is no influence from the boundary.

The spatial operators $\partial_r$ are discretized using a fifth-order accurate finite difference operator $D_{6-5}$ satisfying the summation by parts property and the no-incoming boundary condition is implemented through a penalty method. The time derivatives $\partial_t$ are discretized using a standard fourth-order Runge-Kutta algorithm. For more details on the definition of the $D_{6-5}$ operator, the penalty method and numerical time-integrators we refer the reader to Ref.~\cite{oSmT12} and references therein.

We have tested our code for the Regge-Wheeler equation on a Schwarzschild background metric in ingoing Eddington-Finkelstein coordinates~\cite{oSmT01}, for which $R=r$, $\gamma(r) = 1/\alpha(r) =\sqrt{1 + r_H/r}$, $\beta(r) = r_H/(r\gamma(r)^2)$ and $U_\ell(r) = -3r_H/r^3 + \ell(\ell+1)/r^2$ are substituted into Eqs.~(\ref{Eq:FOSH1},\ref{Eq:FOSH2},\ref{Eq:FOSH3}). We checked fifth-order self-convergence of the field $\phi$, and by measuring the wave forms seen by a static observer at $r = 20r_H$ we reproduced the following quasi-normal frequencies: $s\cdot r_H=-0.178 + 0.747i$ for $\ell=2$, $s~r_H=-0.18541 + 1.19889i$ for $\ell=3$, and $s~r_H=-0.1883 + 1.61836i$ for $\ell=4$, which agree with those given in the literature, see for example table~2 in Ref.~\cite{hN99}.

\subsection{Wave forms for a static observer}

\begin{figure}[ht]
\resizebox{8cm}{!}{\includegraphics{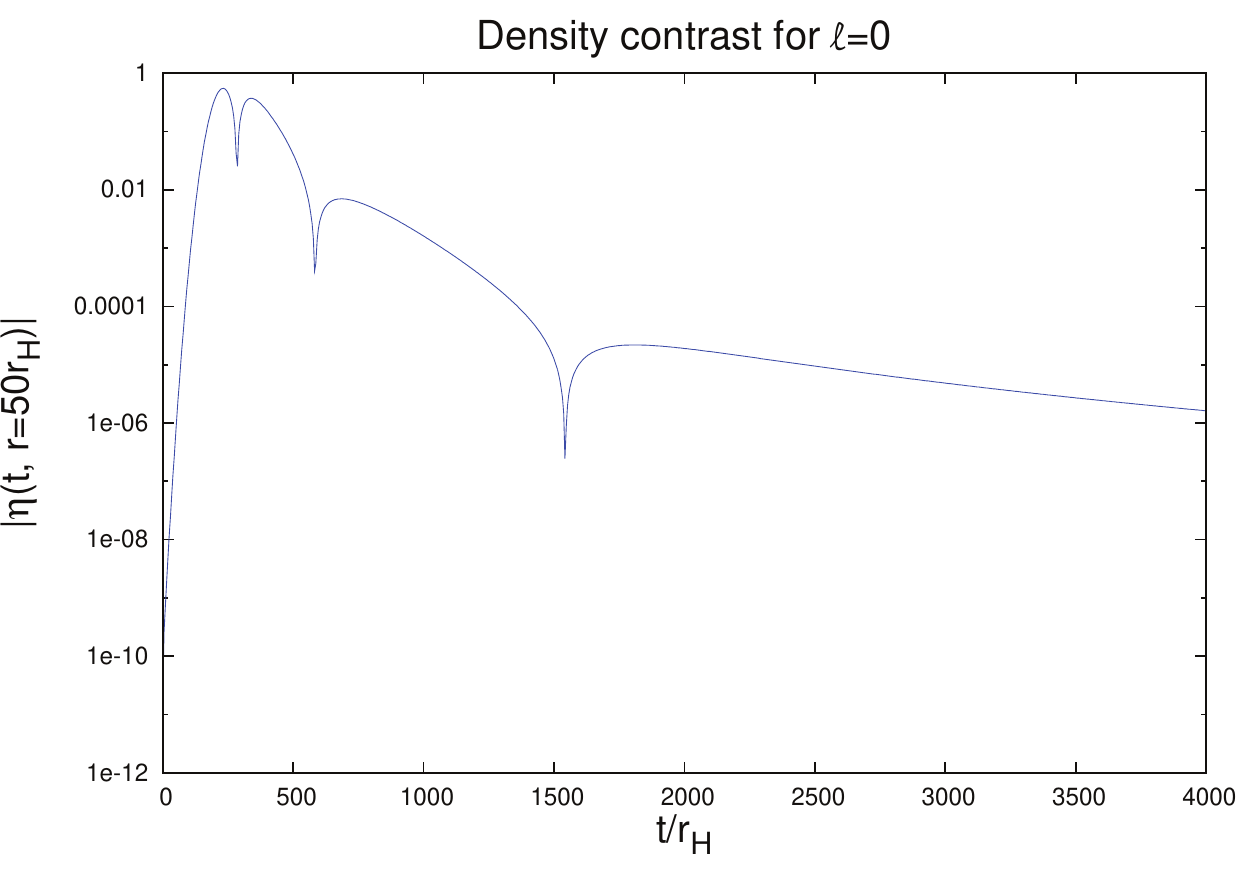}}
\resizebox{8cm}{!}{\includegraphics{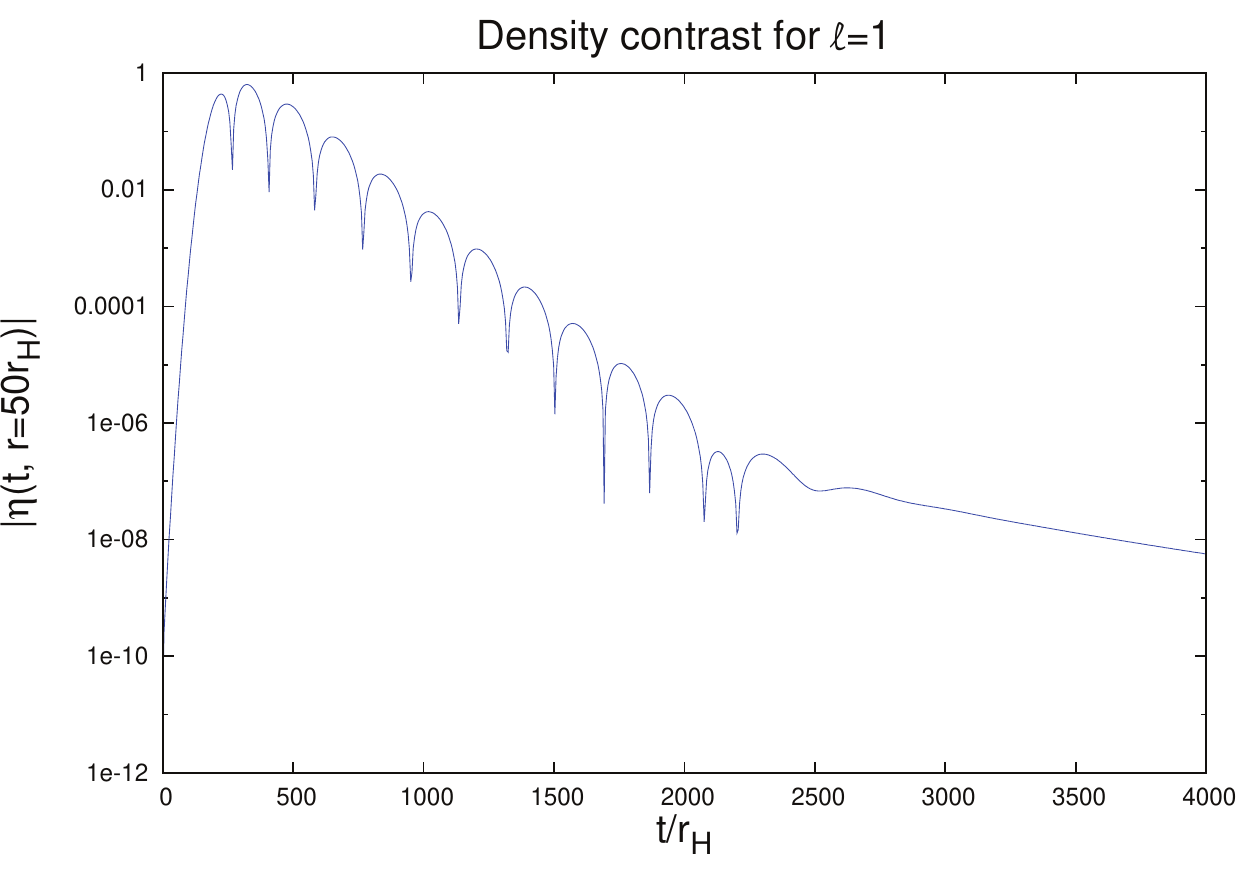}}
\resizebox{8cm}{!}{\includegraphics{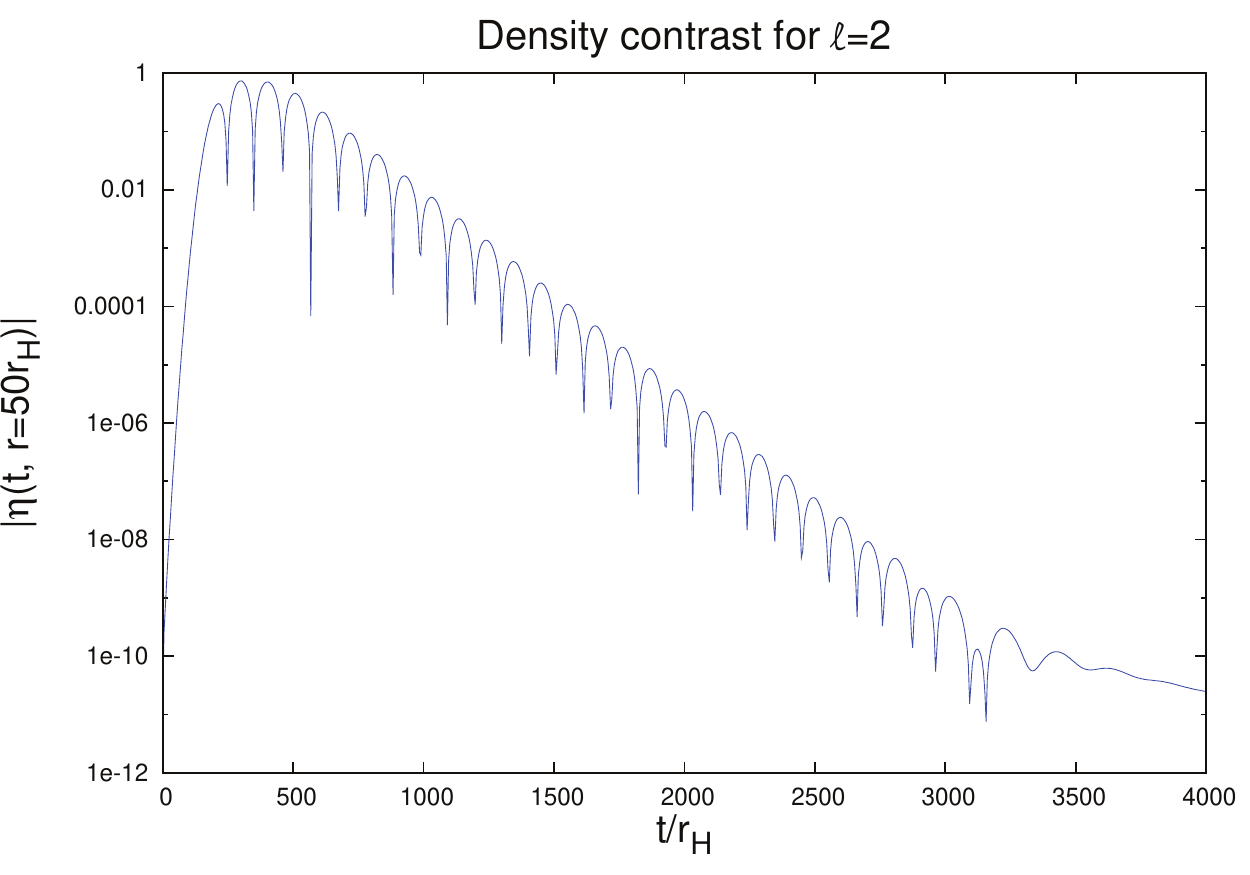}}
\resizebox{8cm}{!}{\includegraphics{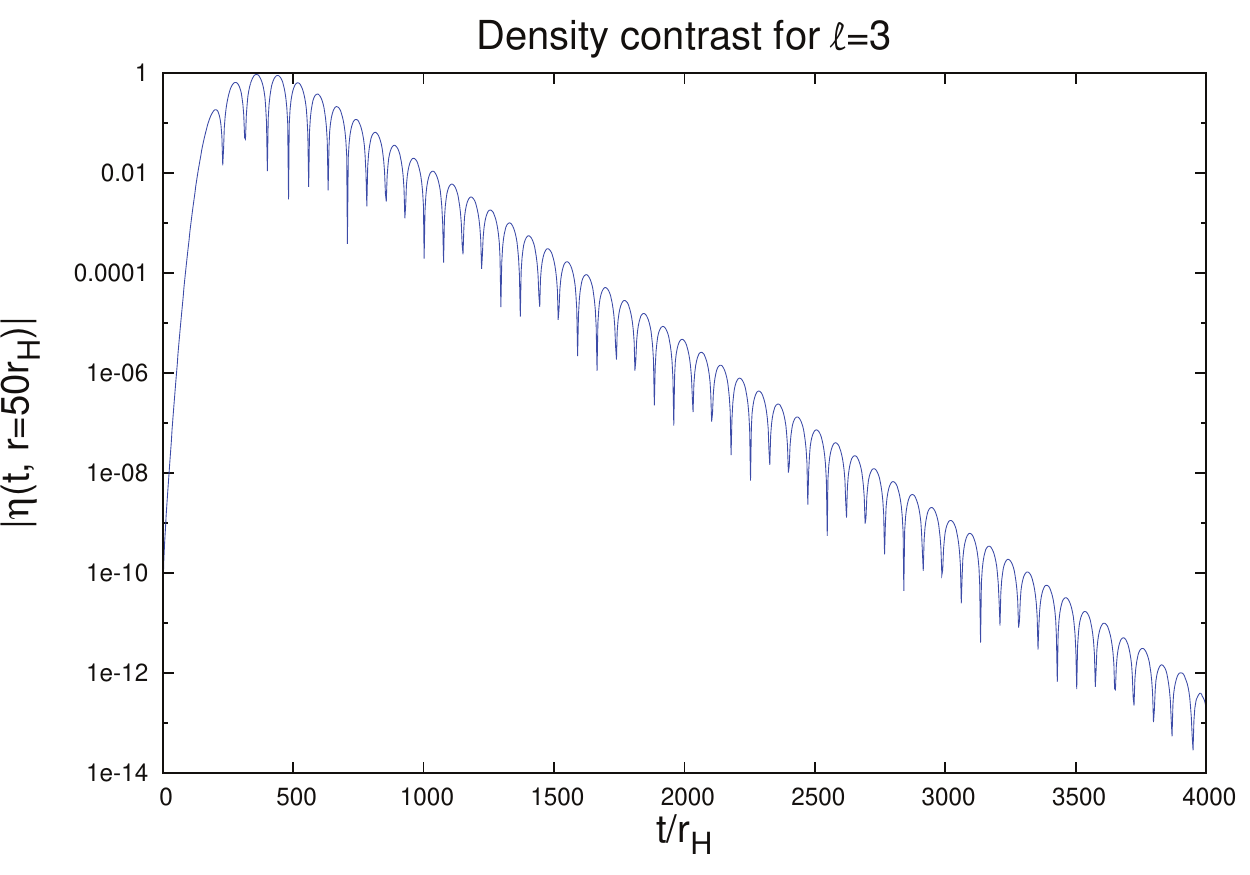}}
\resizebox{8cm}{!}{\includegraphics{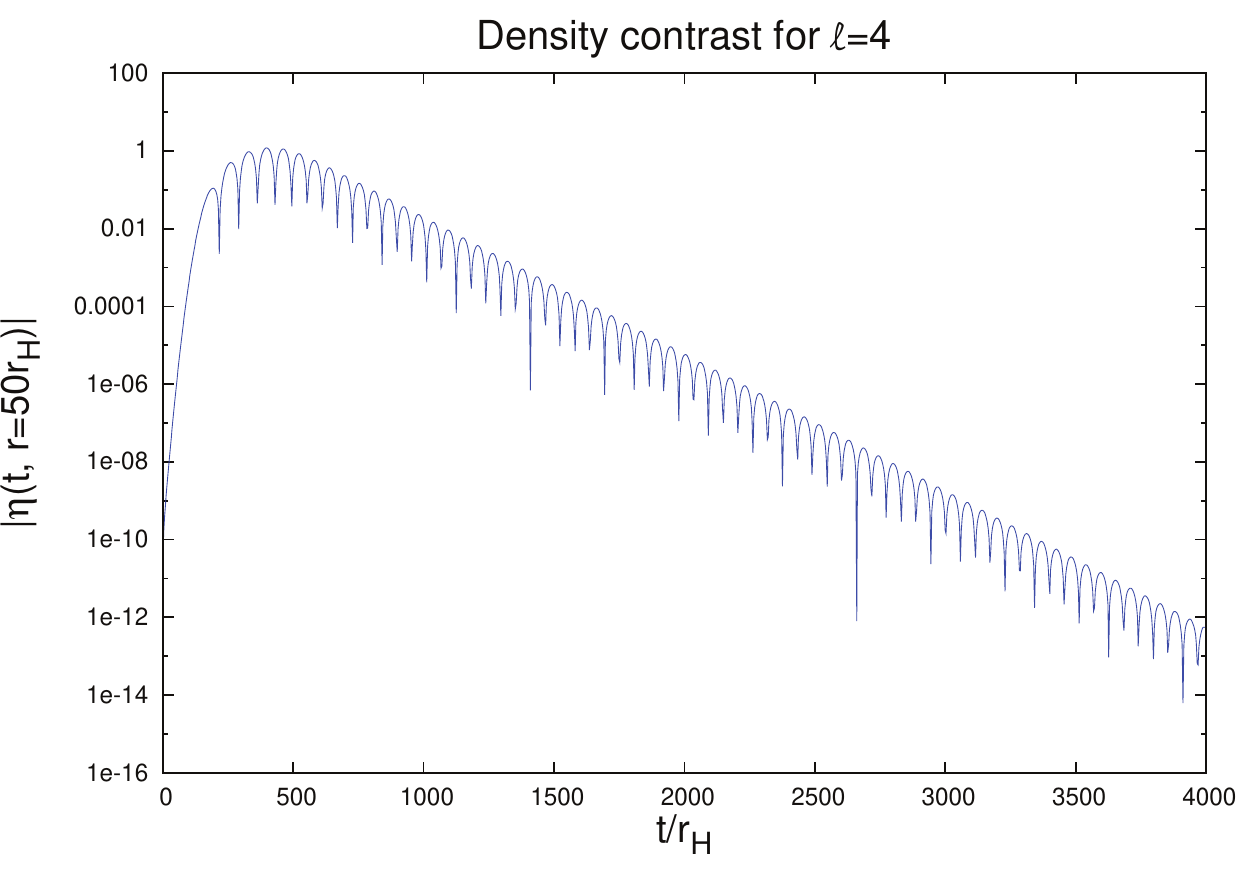}}
\resizebox{8cm}{!}{\includegraphics{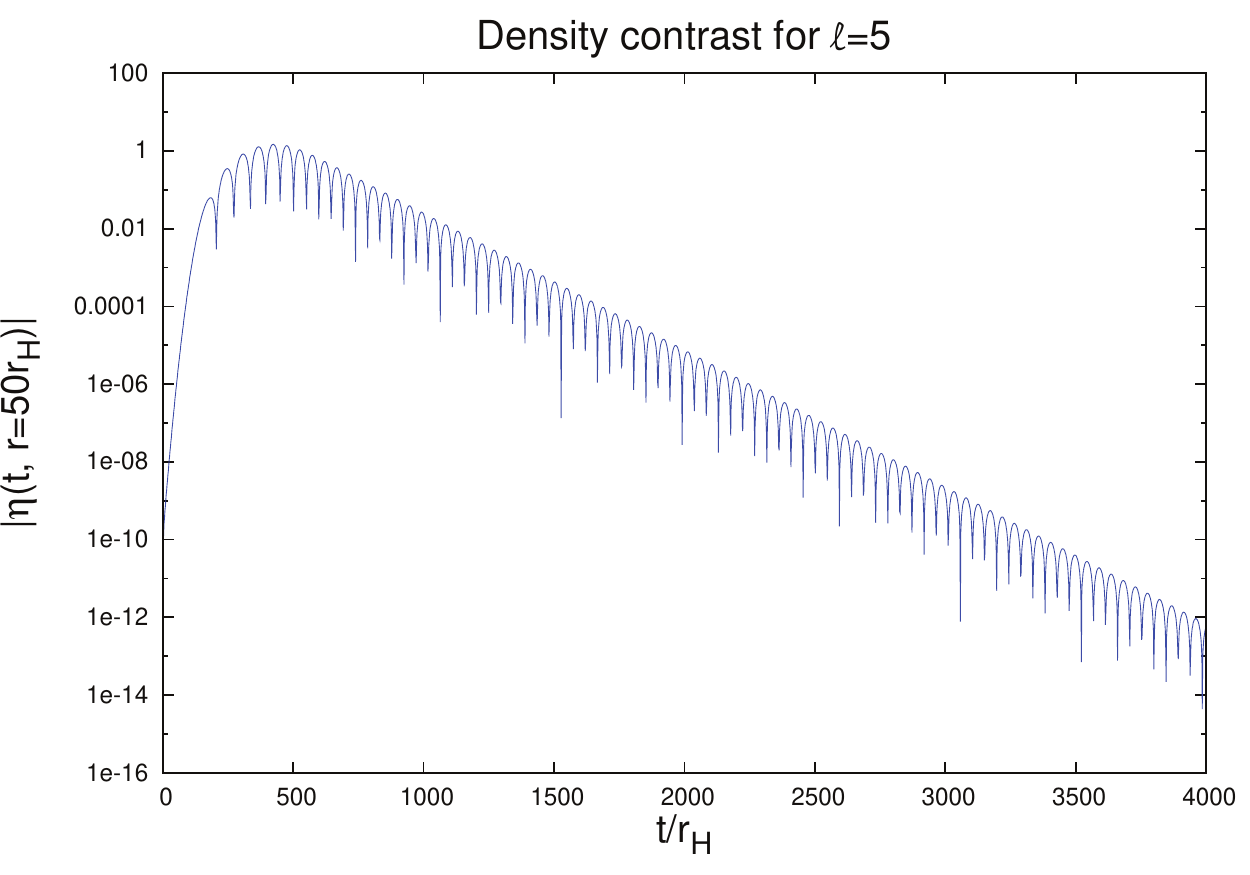}}
\caption{The density contrast parameter $\eta$ vs. time $t$ measured by a static observer at $r = 50r_H$ for different values of the angular momentum parameter $\ell$. In all plots, the sonic horizon is located at $r_c = 7r_H$. Note that only a few oscillations appear in the monopolar case $\ell=0$, preventing us from reading off the quasi-normal frequency in this case. For the cases $\ell=0,1,2$ a late time power-law decay is also visible.}
\label{Fig:decay}
\end{figure}

In Fig.~\ref{Fig:decay} we show the time evolution of the acoustic perturbations measured by a static observer located at $r = 50r_H$ outside the sonic horizon at $r_c = 7r_H$.
The initial data for the evolution consists of a Gaussian pulse with zero initial velocity,
\begin{equation}
f(r) = A\exp\left[ -\frac{1}{2}\left( \frac{r - r_0}{w} \right)^2 \right],
\label{Eq:f}
\end{equation}
with amplitude $A = 1.5$, width $w = 5.0r_H$ and centered at $r_0 = 15r_H$, and
\begin{equation}
\phi(0,r) = f(r),\quad
\chi(0,r) = \frac{1}{\gamma(r)} f'(r),\quad
\pi(0,r ) = -\frac{\beta(r)}{\alpha(r)} f'(r).
\label{Eq:ID}
\end{equation}
In our simulations, we placed the outer boundary at $r=1300r_H$ and used $2^k\times 4000$ grid points, where we varied $k$ over $0,1,2,3,4$ in order to perform convergence tests. We used a Courant factor of $0.5$. The background fluid describing the Michel flow is a polytrope with adiabatic index $\gamma = 1.3333$. The quantities shown in the plots of Fig.~\ref{Fig:decay} are the multipolar components of the \textit{density contrast}, defined as
\begin{equation}
 \frac{\delta n}{n} = \frac{1}{v_s^2} \frac{\delta h}{h}
 = \frac{1}{r}\sum\limits_{\ell m} \eta_{\ell m} Y^{\ell m},\qquad
\eta_{\ell m} := -\frac{1}{v_s^2 h}\left(\frac{1}{N} \sqrt{N+(u^r)^2} \partial_t\phi_{\ell m}
   + u^r\partial_r\phi_{\ell m} - \frac{1}{r} u^r \phi_{\ell m} \right), \label{Eq:DContr}
\end{equation}
where we use Eq.~(\ref{Eq:FOSH1}) and the definition of the auxiliary field $\chi$ in order to rewrite $\partial_t\phi_{\ell m} = \alpha \pi + \beta\gamma\chi$ and $\partial_r\phi_{\ell m} = \gamma \chi$, respectively. As is apparent from these plots, there is an initial burst of radiation which is followed by several cycles of oscillations. The plots corresponding to the cases $\ell = 0,1,2$ show that these oscillations are taken over by a power-law decay at late times. For the remaining cases $\ell > 2$ this is probably also true; however, obtaining the power-law tail would require much higher resolution in this case.

For each $\ell > 0$, there is a clear ringdown signal, and we determined the frequency and decay rate of the corresponding fundamental quasi-normal oscillations by fitting the numerical data to the function
$$
C e^{\sigma t}\sin(\omega t - \delta) 
$$
with free parameters $C$, $\sigma$, $\omega$ and $\delta$. The fit is performed in a time window $[t_1,t_2]$ where the quasi-normal ringing is apparent. The resulting frequencies $s = \sigma + i\omega$ are shown in table~\ref{Tab:QNMCauchy}. Since there is no clear ringdown signal for the particular case $\ell = 0$, only results for $\ell > 0$ are shown. The number of significant figures shown has been estimated by varying the time window and by comparing the results from different resolutions.

\begin{table}[h]
\center
\begin{tabular}{|c|c|c|c|c|}\hline
 $\ell = 1$ & $\ell = 2$ & $\ell = 3$ & $\ell = 4$ & $\ell = 5$ \\
\hline
$-0.0080 + 0.0170i$ & $-0.0081 + 0.0301i$ & $-0.0081 + 0.04271i$ & 
$-0.0081 + 0.0552i$ & $-0.00811 + 0.0677i$\\
\hline
\end{tabular}
\caption{The quasi-normal fundamental frequencies for $r_c = 7r_H$ and $\ell = 1,2,3,4,5$ obtained from the data shown in Fig.~\ref{Fig:decay}.}
\label{Tab:QNMCauchy}
\end{table}

As mentioned above, the late time behavior is characterized by a power-law decay, $\eta \sim t^{-p}$, as is apparent from the plots in Fig.~\ref{Fig:decay} for $\ell = 0,1,2$. We have determined the power $p$, again using a standard fitting routine, obtaining the following results\footnote{See Ref.~\cite{Chingetal94} for a general discussion on the late-time tail decay for wave propagation on a curved spacetime, where it is shown that for a certain class of problems the decay only depends on the asymptotic properties of the effective potential. Since in our case the effective potential ${\cal N} V_\ell$ as a function of the tortoise coordinate $r_*$ decays as $v_\infty^2[ \ell(\ell+1)/r_{*}^2 + C\log(r_*)/r_*^3]$  for large $r_*$ with $C$ a nonvanishing constant, it follows from the results in Ref.~\cite{Chingetal94} that the fluid potential $\Psi$ should decay as $t^{-(2\ell + 3)}$ for $\ell\geq 1$. We have verified that the function $\Psi$ in our simulations reproduce this decay rate for $\ell=0,1,2$ to high accuracy. However, the results in~\cite{Chingetal94} do not apply directly to the density contrast, which is a nontrivial linear combination of $\Psi$ and its first derivatives.}: $p = 3.89 \pm 0.03$ for $\ell = 0$, $p = 6.62 \pm 0.37$ for $\ell = 1$, and $p = 9.16 \pm 0.24$ for $\ell = 2$. The error has been estimated by performing the fit in different time windows lying between $3000r_H$ and $8000r_H$ and by using different resolutions.

\begin{figure}[ht]
\resizebox{11cm}{!}{\includegraphics{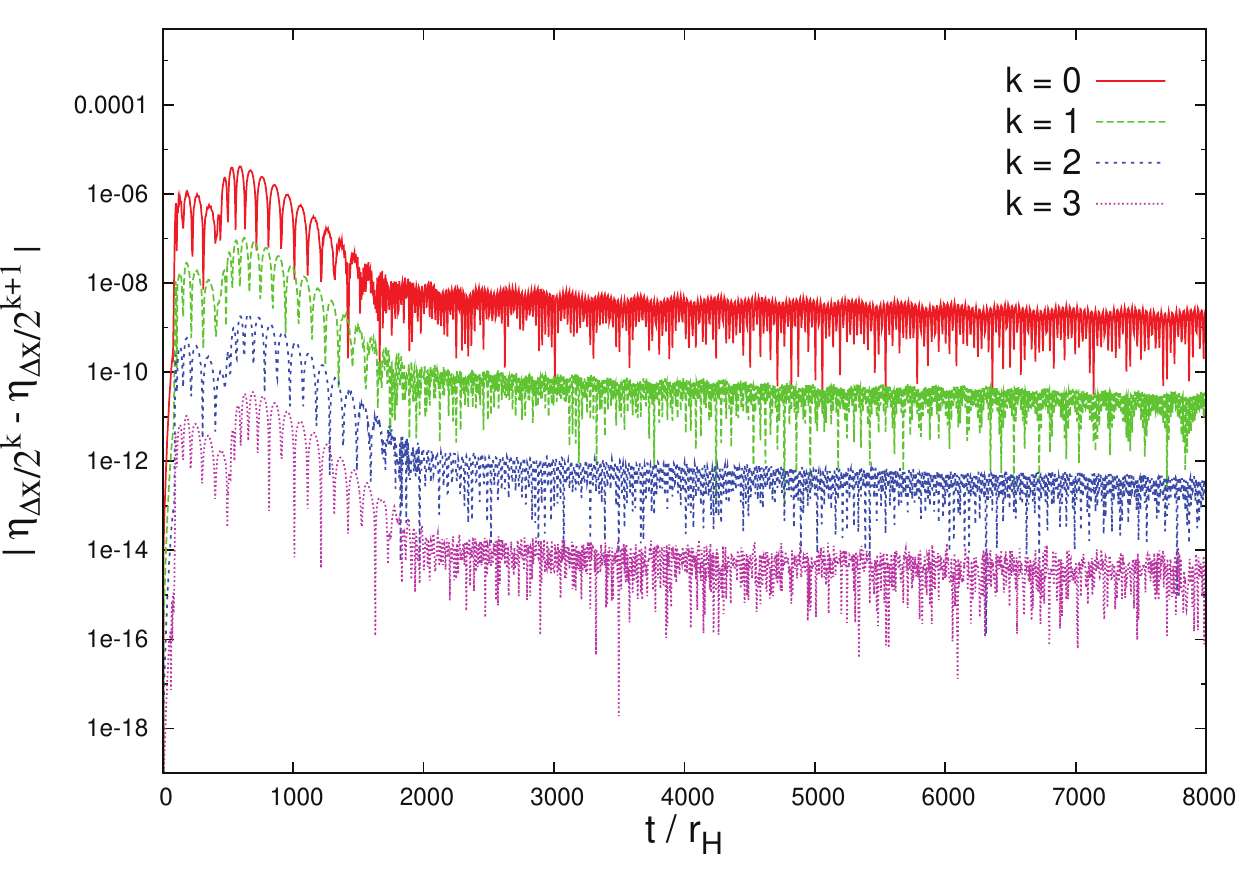}}
\caption{Self-convergence test for the case $\ell=2$ in Fig.~\ref{Fig:decay}. The top curve corresponds to the error between the results using $2^0\times 4000$ and $2^1\times 4000$ grid points, the second curve to the error using $2^1\times 4000$ and $2^2\times 4000$ points, etc. The convergence factor has been estimated to lie close to $5$, indicating fifth-order self-convergence.}
\label{Fig:Convergence}
\end{figure}

In order to check the validity of our numerical results we have performed several self-convergence tests. In Fig.~\ref{Fig:Convergence} we show a particular example in which we plot the difference of the density contrast function $\eta_{\ell m}$ between two consecutive resolutions. This plot corresponds to the quadrupolar case $\ell=2$ in Fig.~\ref{Fig:decay}. Note that there is high frequency noise appearing at $t \sim 1500r_H$, which is probably due to the presence of the time and space derivatives of $\phi$ in the expression for $\eta_{\ell m}$ in Eq.~(\ref{Eq:DContr}). However, it is clear from the plot that the error decreases with increasing resolution. We have estimated the convergence factor to lie close to five, indicating fifth-order self-convergence.

\section{Results for the quasi-normal frequencies}
\label{Sec:Results}

In this section, we present and analyze the results from our calculations of the quasi-normal acoustic frequencies as a function of the sonic radius $r_c$ and the angular momentum $\ell$. All the calculations in this section refer to the Michel flow on a Schwarzschild background for a polytropic fluid with adiabatic index $\gamma = 1.3333$. In Sec.~\ref{SubSec:FF}, we discuss the fundamental frequencies for values of $r_c$ ranging in the interval $[2r_H,30r_H]$ and $\ell = 0,1,\ldots,7$. In Sec.~\ref{SubSec:OT}, we also discuss quasi-normal frequencies corresponding to the first few overtones.

\subsection{Fundamental frequencies}
\label{SubSec:FF}

In table~\ref{Tab:QNM} we show the fundamental monopolar, dipolar and quadrupolar quasi-normal frequencies for different values of $r_c$. These frequencies were calculated using the matching method described in Sec.~\ref{Sec:QNMMode}, and in the dipolar and quadrupolar cases with $r_c/r_H = 2,7,10,20,30$ also using the numerical Cauchy evolution described in the previous section. As can be seen from this table, the two approaches give
results which are consistent within their numerical errors.

\begin{table}[h]
\center
\begin{tabular}{|r|c||c|c|c|c|}\hline
$r_c/r_H$ & $\kappa\cdot r_H$ & 
$s\cdot r_H$ $(\ell = 0)$ & $s\cdot r_H$ $(\ell = 1)$ & $s\cdot r_H$ $(\ell = 2)$\\
\hline
$2$ & $0.16536$ & 
$-0.05947+0.02661i$ &  $-0.06174+0.1398i$ & $-0.06203+0.2416i$\\
& & & ($-0.06+0.14i$) & ($-0.062+0.242i$)
\\ 
$3$ & $0.08334$ & 
$-0.02932+0.008119i$ & $-0.03144+0.06813i$ & $-0.03162+0.1191i$
\\
$4$ & $0.05182$ & 
$-0.01805 + 0.003508i$ & $-0.01965 + 0.04204i$ & $-0.01977 + 0.07381i$
\\
$5$ & $0.03606$ & 
$-0.01249 + 0.001812i$ & $-0.01372 + 0.02920i$ &  $-0.01380 + 0.05138i$
\\
$6$ & $0.02690$ &
$-0.009286 + 0.001042i$ & $-0.01026 + 0.02178i$ & $-0.01032 + 0.03838i$
\\
$7$ & $0.02104$ & $-0.007250 + 0.0006431i$ 
& $-0.008036 + 0.01704i$ & $-0.008086 + 0.03006i$ \\
 & & & ($-0.0080 + 0.0170i$) & ($-0.0081 + 0.0301i$)
\\
$8$ & $0.01703$ & 
$-0.005858 + 0.0004166i$ & $-0.006513 + 0.01381i$ & $-0.006553 + 0.02436i$
\\
$9$ & $0.01414$ & 
$-0.004861 + 0.0002792i$ & $-0.005416 + 0.01148i$ & $-0.005449 + 0.02026i$
\\
$10$ & $0.01199$ & $-0.004118 + 0.0001913i$ & 
$-0.004595 + 0.009736i$ & $-0.004623 + 0.01720i$ \\
 & & & ($-0.0046 + 0.0097i$) & ($-0.00462 + 0.0172i$)
\\
$20$ & $0.00410$ & & $-0.001577 + 0.003344i$ & $-0.001586 + 0.005914i$ \\
 & & & ($-0.0016 + 0.0033i$) & ($-0.0016 + 0.00591i$)
\\
$30$ & $0.00220$ &  & $-0.0008493 + 0.001803i$ & $-0.0008546 + 0.003190i$ \\
 & & & ($-0.00085 + 0.0018i$) & ($-0.00085 + 0.00319i$)
\\
\hline
\end{tabular}
\caption{Fundamental quasi-normal frequencies for acoustic perturbations of the Michel flow for different values of $r_c$ and $\ell$. The frequencies in the first line of each entry for $r_c/r_H$ are the ones obtained from the matching procedure discussed in Sec.~\ref{Sec:QNMMode}, and four significant figures are shown. The frequencies in parenthesis refer to the ones obtained from the Cauchy evolution code and are shown for comparison. In the monopolar case $\ell=0$ we have not been able to obtain the frequencies from the Cauchy evolution, for the reasons described in the previous section. For $\ell=0$ and $r_c > 15r_H$ we have not been able to compute the frequencies in a reliable way using our matching procedure; their computation seems to require higher accuracy than the one available in our current code.}
\label{Tab:QNM}
\end{table}

\begin{figure}[ht]
\resizebox{8.5cm}{!}{\includegraphics{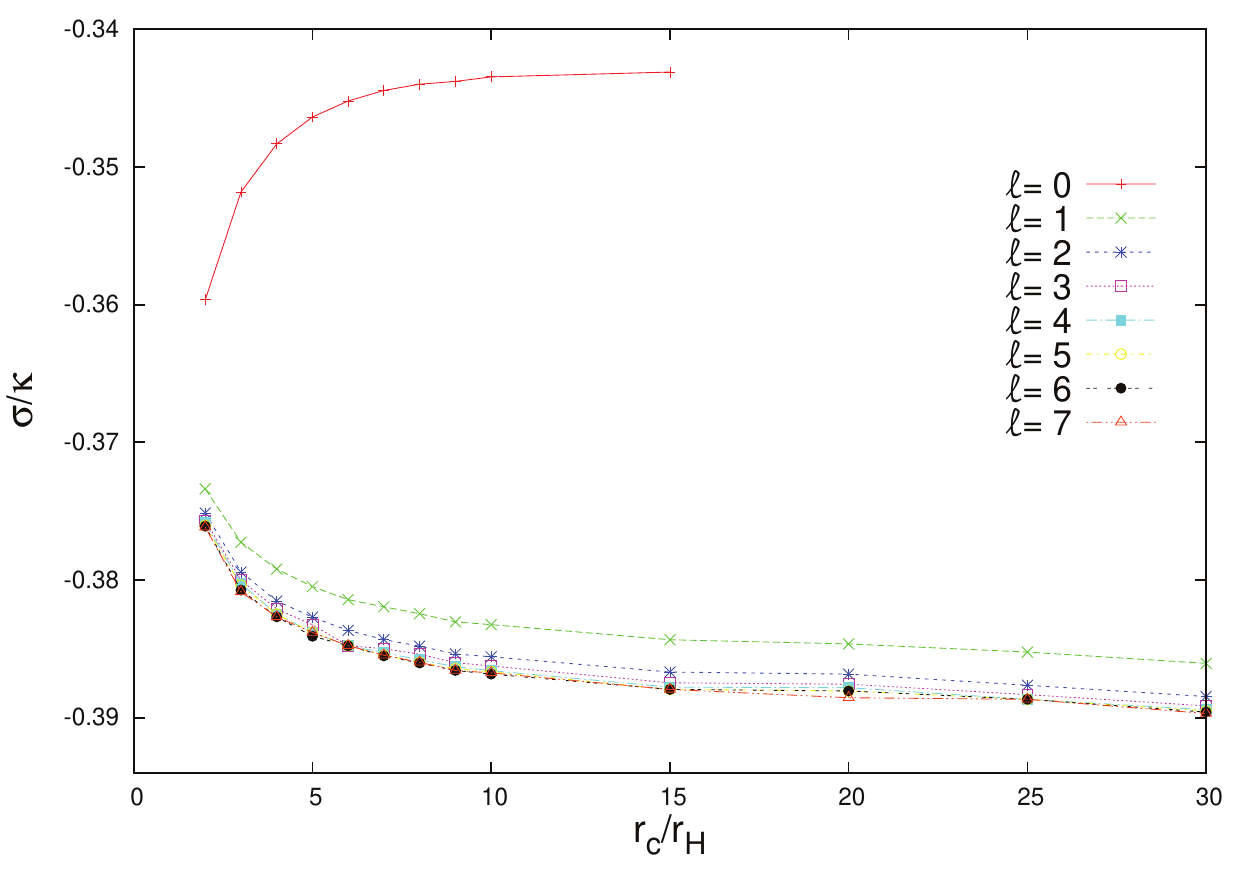}}
\resizebox{8.5cm}{!}{\includegraphics{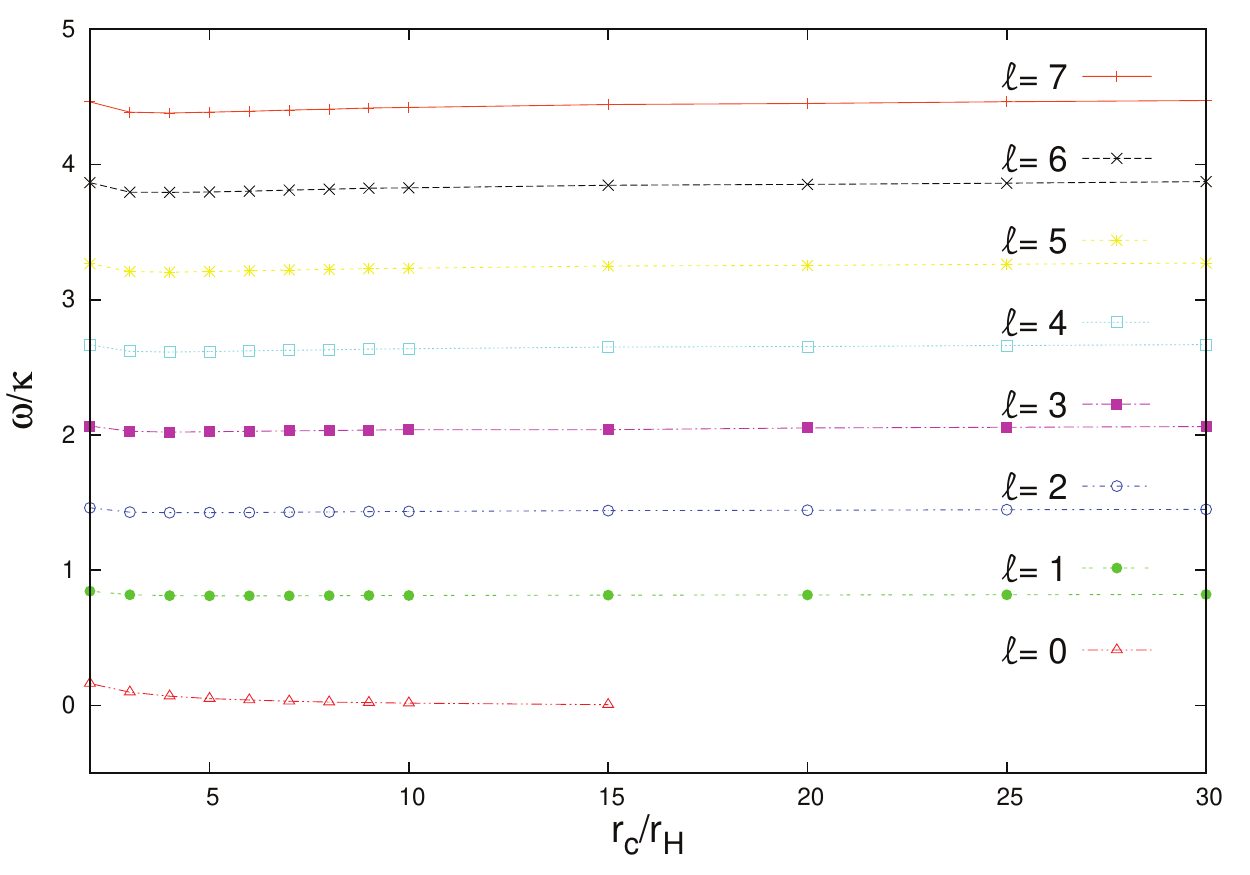}}
\caption{The fundamental quasi-normal frequencies in units of $\kappa$ as a function of $r_c$. Left panel: real part $\sigma/\kappa$ divided by the surface gravity $\kappa$ for $\ell = 0,1,2,\ldots,7$. As is apparent from the plot, for $\ell \neq 0$ and $r_c/r_H \geq 10$ these values are almost independent of $\ell$ and $r_c$, and can be approximated by $-0.387$ to about $1\%$ accuracy. Right panel: imaginary part $\omega/\kappa$ divided by the surface gravity for $\ell = 0,1,2,\ldots,7$. These values are almost independent of $r_c$, and we found that for $\ell\neq 0$ they are well-approximated by the empiric formula $0.21 + 0.606\ell$.}
\label{Fig:QNM}
\end{figure}

Also shown in table~\ref{Tab:QNM} are the values for the surface gravity $\kappa$ of the acoustic metric, computed using Eq.~(\ref{Eq:SurfaceGravity}). It turns out $\kappa$ plays an important role for understanding the behavior of the quasi-normal frequencies as a function of the location of the sonic horizon $r_c$. Indeed, $\kappa$ has units of frequency (in geometrized units) and thus it is natural to analyze the quasi-normal frequencies in units of $\kappa$. In Fig.~\ref{Fig:QNM} we show plots of $s/\kappa$ vs. $r_c$ for the fundamental quasi-normal frequencies $s = \sigma + i\omega$ for different values of $\ell$. As is apparent from these plots, the value of $s/\kappa$ seems to be almost independent of $r_c$ for large $r_c/r_H$. Specifically, we have found that the empiric formula
\begin{equation}
\frac{s}{\kappa} \simeq -0.387 + (0.21 + 0.606\ell)i,\qquad
10 \leq \frac{r_c}{r_H}\leq 30,\quad \ell = 1,2,\ldots,7,
\label{Eq:Empirical}
\end{equation}
gives a fit for the fundamental frequency to a relative accuracy better than $2\%$. Notice that in the monopolar case $\ell=0$ the behavior of $\sigma$ as a function of $r_c$ is different than for higher multipoles $\ell\geq 1$.

\subsection{Overtones}
\label{SubSec:OT}

Using the matching procedure described in Sec.~\ref{Sec:QNMMode} we have also computed the quasi-normal frequencies of the first few excited modes. In table~\ref{Tab:QNMExcited} we present two examples for the quasi-normal spectrum, referring to dipolar and quadrupolar acoustic perturbations, respectively, with $r_c = 10r_H$.

\begin{table}[h]
\center
\begin{tabular}{|r||c|c||}\hline
$n_o$ & $s\cdot r_H (\ell = 1)$ & $s\cdot r_H (\ell = 2)$ \\
\hline
$1$ & $-0.01503 + 0.007955i$ & $-0.01437 + 0.01590i$ \\
$2$ & $-0.02691 + 0.006537i$ & $-0.02526 + 0.01408i$ \\
$3$ & $-0.03907 + 0.005732i$ & $-0.03699 + 0.01257i$ \\
$4$ & $-0.05123 + 0.005219i$ & $-0.04905 + 0.01149i$ \\
$5$ & $-0.06336 + 0.004855i$ & $-0.06120 + 0.01071i$ \\
$6$ &                                         & $-0.07336 + 0.01012i$ \\
$7$ &                                         & $-0.08551 + 0.009659i$ \\
$8$ &                                         & $-0.09764 + 0.009281i$ \\
$9$ &                                         & $-0.1098 + 0.008965i$ \\
\hline
\end{tabular}
\caption{Quasi-normal dipolar and quadrupolar frequency spectrum for acoustic perturbations of the Michel flow with sonic horizon located at $r_c = 10r_H$. $n_o = 1$ denotes the first overtone, $n_o = 2$ the second etc. Four significant figures are shown. Modes with excitation numbers $n_0 > 5$ for $\ell = 1$ and $n_0 > 9$ for $\ell = 2$ could not be obtained in a reliable way with the current version of our code, since the computation of their frequency seems to require a more powerful Newton algorithm or higher accuracy.}
\label{Tab:QNMExcited}
\end{table}

\begin{figure}[ht]
\resizebox{8.5cm}{!}{\includegraphics{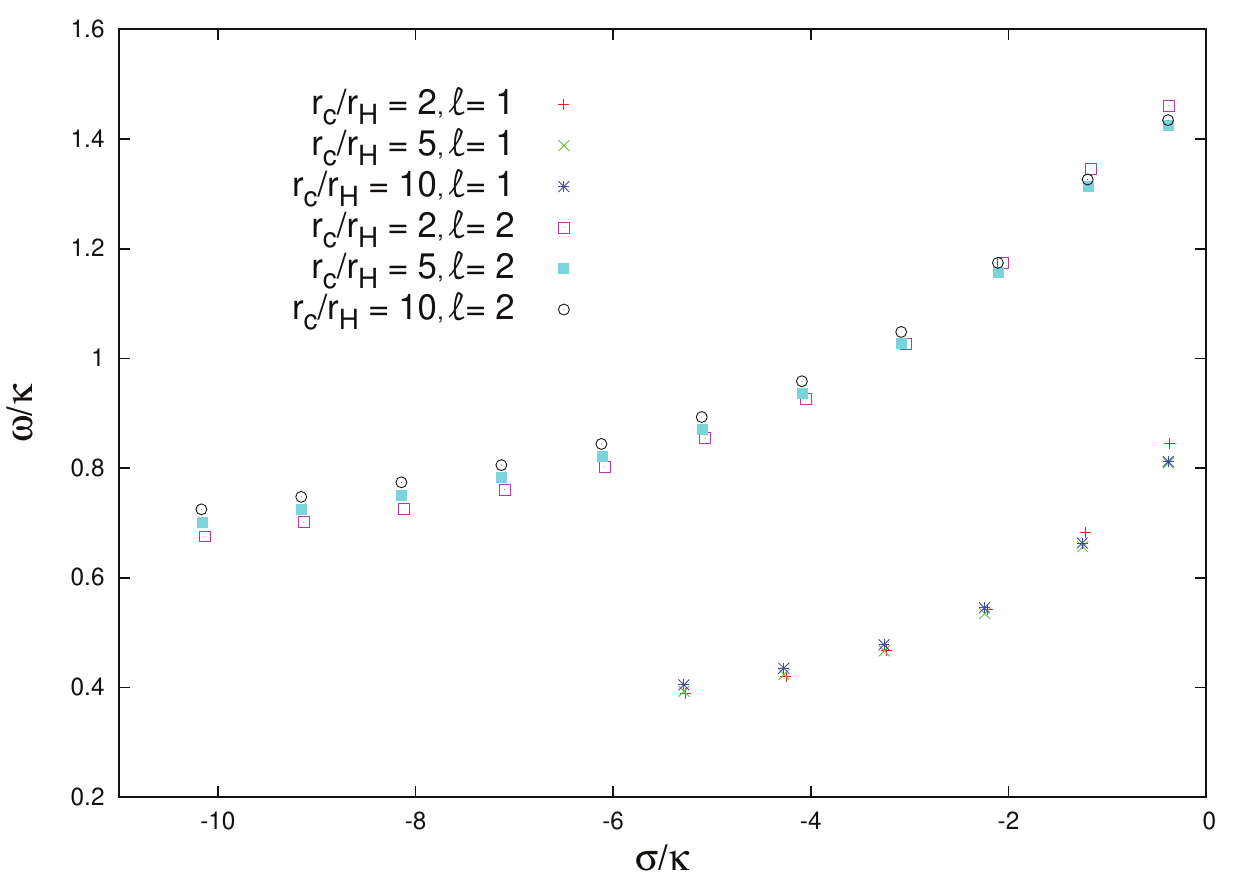}}
\caption{The spectrum of the quasi-normal acoustic excitations. Shown are the imaginary vs. the real part of the frequencies $s/\kappa$  divided by the surface gravity for $r_c/r_H = 2,5,10$ and $\ell=1,2$. As is apparent from the plot, the spectrum is approximately independent of $r_c$.}
\label{Fig:QNMExcited}
\end{figure}

In Fig.~\ref{Fig:QNMExcited} we show plots of the quasi-normal dipolar and quadrupolar spectrum in units of the surface gravity $\kappa$ for different values of $r_c$. As in the case of the fundamental frequencies, we appreciate from these plots that the spectrum of quasi-normal excitations seems to be nearly independent of $r_c$, indicating that the frequencies scale like $\kappa$.

\subsection{Eikonal limit}
\label{SubSec:Eikonal}

In the high-frequency limit, the quasi-normal oscillations can be interpreted in terms of wave packets which are concentrated along a circular null geodesics and decay because the circular null geodesic is unstable, see Refs.~\cite{vCaMeBhWvZ09,eBvCaS09} and references therein for more details. Therefore, one expects that in this limit the quasi-normal frequencies $s$ are related to the properties of the unstable circular null geodesics. As shown in~\cite{vCaMeBhWvZ09} the imaginary part $\omega = \im(s)$ of $s$, describing the oscillatory behavior, is directly related to the angular velocity of the unstable circular null geodesic, while the real part $\sigma = \re(s)$ is equal to its Lyapunov exponent.

As can be deduced from the analysis in~\cite{vCaMeBhWvZ09} an arbitrary asymptotically flat, static spherically symmetric metric of the form
\begin{equation}
ds^2 = -f(r) dT^2 + \frac{dr^2}{g(r)} 
 + r^2\left( d\vartheta^2 + \sin^2\vartheta d\varphi^2 \right)
\label{Eq:SphSymMetric}
\end{equation}
with time coordinate $T$ and positive smooth functions $f(r)$ and $g(r)$ possesses an unstable circular null geodesic at $r = r_{circ}$ if and only if the function $H(r) := f(r)/r^2$ has a local maximum at $r = r_{circ}$, and in this case the associated angular velocity and Lyapunov exponent are given by
\begin{equation}
\Omega_{circ} = \sqrt{H(r_{circ})},\qquad
\lambda_{circ} = \left. \sqrt{\frac{f(r) g(r)}{2}}
 \sqrt{ -\frac{1}{H(r)} \frac{d^2}{dr^2} H(r) } \right|_{r = r_{circ}}.
\end{equation}
For high values of $\ell$, these parameters determine the quasi-normal frequencies according to the formula
\begin{equation}
s = -(n_o + 1/2)\lambda_{circ} + i\ell\,\Omega_{circ},
\end{equation}
with $n_o$ the overtone number, see~\cite{vCaMeBhWvZ09}. Comparing Eq.~(\ref{Eq:SphSymMetric}) with the form~(\ref{Eq:SoundMetricMichelDiag}) of the acoustic metric and discarding the conformal factor $n/(h v_s)$ which does not affect the null geodesics as trajectories in spacetime, we find that in our case $f(r) = X(r) v_s^2$ and $g(r) = X(r)$, such that
$$
H(r) = \frac{X(r) v_s^2}{r^2},\qquad \sqrt{f(r) g(r)} = X(r) v_s.
$$
\begin{figure}[ht]
\resizebox{8.5cm}{!}{\includegraphics{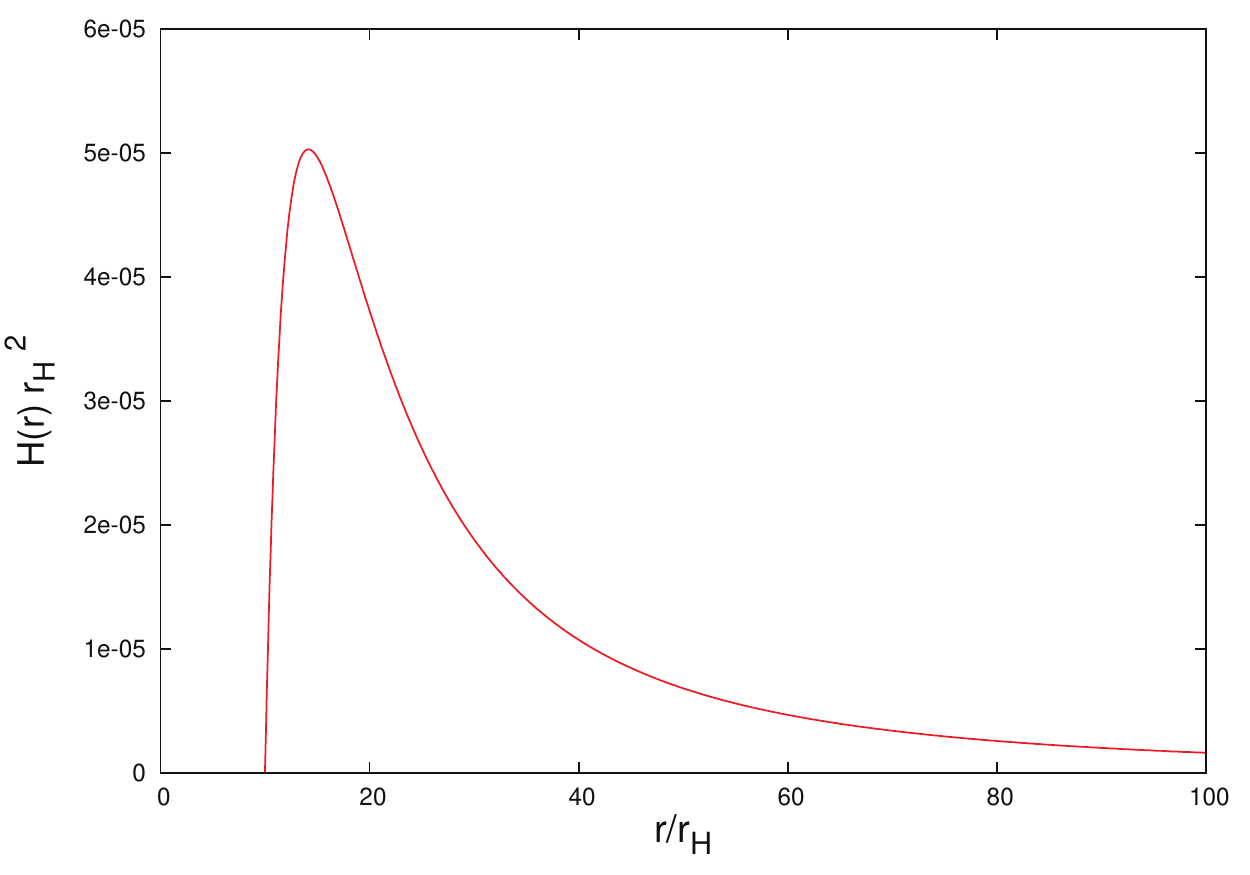}}
\caption{Graph of the function $H(r)$ for the case where the sonic horizon is located at $r_c/r_H = 10$. As is clearly visible from the plot, this function has a maximum where the acoustic metric has an unstable circular null geodesics. This maximum is numerically determined to be located at $r_{circ}/r_H = 14.158$.}
\label{Fig:FunctionH}
\end{figure}
A plot of the function $H(r)$ for the case $r_c/r_H = 10$ is given in Fig.~\ref{Fig:FunctionH}, which shows that the acoustic metric for the Michel flow admits unstable circular null geodesics. Numerically, we find the values $r_{circ}/r_H = 14.158$, $\Omega_{circ}/\kappa = 0.59159$ and $\lambda_{circ}/(2\kappa) = 0.38637$ which agree remarkably well with the corresponding values in the empirical formula~(\ref{Eq:Empirical}) describing the fundamental frequencies. We have repeated the analysis for higher values of $r_c/r_H$ ranging between $10$ and $30$, finding similar values for $\Omega_{circ}/\kappa$ and $\lambda_{circ}/(2\kappa)$ (the difference is less than $1\%$).

\section{Conclusions}
\label{Sec:Conclusions}

In this work, we have analyzed spherical and nonspherical acoustic perturbations of the Michel flow, which describes a perfect fluid which falls radially into a Schwarzschild black hole. As shown by Moncrief~\cite{vM80}, the equations of motion for such perturbations can be cast into a wave equation on a curved effective background geometry described by the acoustic metric. For the case of the Michel flow, the acoustic metric has the same qualitative properties as a black hole spacetime and thus describes a natural analogue black hole.

Using this natural astrophysical analogue black hole, we have shown by numerical computation that when perturbed, the Michel flow exhibits quasi-normal acoustic oscillations. We have computed the associated frequencies $s = \sigma + i\omega$ using two different methods. The first method which, to our knowledge, is new is based on matching the two local solutions $\psi_+(s,r)$ and $\psi_-(s,r)$ of the radial mode equation which, for $\re(s) > 0$, are decaying as $r\to \infty$ and $r\to r_c$, respectively. A common challenge for computing the quasi-normal modes is to determine the analytic continuation of these functions for $\re(s) < 0$ and to find the complex frequencies $s$ for which $\psi_+$ and $\psi_-$ are linearly dependent. While in some cases the solutions $\psi_\pm$ can be represented by simple series expansions and the quasi-normal frequencies can be found using continued fraction techniques~\cite{MQNdeS}, in our problem the effective potential appearing in the mode equation is not even known in closed form and so more general methods are required. The new ingredient of our method consists in computing the analytic continuations of $\psi_\pm$ via a Banach iteration technique, where each iteration leads to an improved approximation for the solution. Each iteration involves computing a line integral in the complex $r$-plane which converges for all $\omega = \im(s) > 0$. While the integral in each iteration needs to be computed accurately, we have found that only a few iterations are needed in order to achieve high accuracy. The two solutions $\psi_\pm$ are then matched by finding the zeros of their Wronski determinant using a standard Newton algorithm.

Our method is rather general and does not depend on the details of the effective potential except for the fact that it should possess a sufficiently well-behaved analytic continuation on the complex $r$-plane. What precisely we mean by ``sufficiently well-behaved" will be explained in detail elsewhere~\cite{eCoS15b}, but it seems flexible enough to comprise many relevant effective potentials found in general relativity (including the Regge-Wheeler potential and its generalization to the Reissner-Nordstr\"om case). Another advantage of our method is that it does not require a closed-form expression for the effective potential. For the case of acoustic perturbations of the Michel flow considered in this article the potential is only known in implicit form, though it is analytic in $1/r$ as we have shown in the appendix. In the cases we have analyzed here, our method seems to work very well to find the fundamental frequencies and the first few overtones. However, so far our code fails to find very high overtones. The reason for this is probably related to our simple Newton algorithm and our crude finite-difference approximation for the derivative of Wronski determinant.

Using our method we have computed the quasi-normal acoustic frequencies of the Michel flow for different values $r_c$ and $r_H$ of the sonic and event horizon radii, and for different values of the angular momentum number $\ell$. By means of the Cauchy code described in Sec.~\ref{Sec:QNMCauchy} we have verified the validity of the fundamental frequency for $\ell > 0$, and also computed the late time power-law decay rate in some cases. Although in general the frequency spectrum depends on two parameters $r_H$ and $r_c$, or, equivalently, on $r_H$ and the surface gravity $\kappa$ of the acoustic hole, we found that for $r_c \gg r_H$ the quasi-normal frequencies $s$ scale like $\kappa$, the parameter $r_H$ becoming unimportant. Furthermore, for $r_c\gg r_H$ the real part of $s$ describing the decay rate depends only mildly on $\ell$ for $\ell\geq 1$. Specifically, we have found the following empiric formula for the fundamental frequency:
\begin{equation}
s\kappa^{-1} \simeq -0.387 + (0.21 + 0.606\ell)i
\end{equation}
for $\ell = 1,2,\ldots,7$ and $r_c$ ranging in the interval between $10r_H$ and $30r_H$, with $r_H$ the event horizon radius. In the limit where the sound speed $v_\infty \ll c$ at infinity is much smaller than the speed of light, $\kappa$ can be given by a simple analytic formula. It follows from Eq.~(\ref{Eq:SurfaceGravity}) and standard expansions in $\nu_\infty := v_\infty/c$~\cite{Shapiro-Book,eCoS15a} that
\begin{equation}
\kappa \simeq \frac{8\nu_\infty^3}{r_H} \simeq \frac{1}{2}\sqrt{\frac{r_H}{2r_c^3}},
\end{equation}
for a polytropic equation of state with $\gamma = 4/3$. For a black hole of mass $M$ this gives
\begin{equation}
\kappa \simeq 8\times 10^5 \left( \frac{M_\odot}{M} \right) \frac{\nu_\infty^3}{s},
\end{equation}
with $M_\odot$ the solar mass.

Although in this work we have restricted ourselves to a polytropic fluid with adiabatic index $\gamma = 1.3333\simeq 4/3$, other fluid flows could be analyzed with our method, provided they are described by an analytic equation of state satisfying the assumptions (F1)--(F3) listed in Sec.~\ref{Sec:Michel}. Furthermore, based on our general results in Ref.~\cite{eCoS15a}, it should not be difficult to generalize our calculations to more general nonrotating black holes, and to analyze the dependency of the quasi-normal acoustic frequencies on the background metric. It would be interesting to study the impact of these acoustic oscillations on the emission of electromagnetic and gravitational radiation.


\acknowledgments

It is our pleasure to thank Luis Lehner and Thomas Zannias for fruitful and stimulating discussions. EC and OS thank the gravitational physics group at University of Vienna, where part of this work was performed, for their hospitality. MM and OS also thank the Perimeter Institute for Theoretical Physics for hospitality. This research was supported in part by CONACyT Grants No. 238758, 236810, 233137, by a CIC Grant to Universidad Michoacana and by Perimeter Institute for Theoretical Physics. Research at Perimeter Institute is supported by the Government of Canada through Industry Canada and by the Province of Ontario through the Ministry of Research and Innovation.

\appendix
\section{Analytic continuation of the functions ${\cal N}$ and $V_\ell$}

In this appendix, we prove that the functions ${\cal N}(r)$ and $V_\ell(r)$ in the mode equation~(\ref{Eq:psi}) admit analytic continuations on the domain $\re(r) > r_H$ with the properties that
\begin{equation}
\lim\limits_{\substack{r\to\infty \\ \re(r) > r_H}} {\cal N}(r) = v_\infty,\qquad
\lim\limits_{\substack{r\to\infty \\ \re(r) > r_H}} r^2 V_\ell(r) = v_\infty\ell(\ell + 1),
\label{Eq:Limits}
\end{equation}
where $v_\infty := v_s(n_\infty) > 0$ is the sound speed at infinity. For this, we need to assume that in addition to the properties (F1)--(F3) the specific enthalpy $h(n)$ is an analytic function of $n$. For definiteness, we shall assume that $h(n)$ is given by the polytropic equation of state, Eq.~(\ref{Eq:hn}), which is analytic on the domain $\re(n) > 0$. 

Under these assumptions, we first prove that the Michel flow solution $n(r)$, which is implicitly determined by Eq.~(\ref{Eq:Fundamental}), possesses an analytic continuation on the domain $\re(r) > r_H$ such that
\begin{equation}
\lim\limits_{\substack{r\to\infty \\ \re(r) > r_H}} n(r) = n_\infty,
\label{Eq:Limitn}
\end{equation}
where $n_\infty > 0$ is the particle density at infinity. In order to prove this statement, following~\cite{eCoS15a} we introduce dimensionless quantities $x := r/r_H$, $z := n/n_0$, $n_0 :=(e_0/K)^{1/(\gamma-1)}$, in terms of which Eq.~(\ref{Eq:Fundamental}) can be rewritten as
\begin{equation}
F_\mu(x,z) := f(z)^2\left[ 1 - \frac{1}{x} + \frac{\mu^2}{x^4 z^2} \right] = f_\infty^2 = const.,
\label{Eq:Fmu}
\end{equation}
where $f(z) = 1 + z^{\gamma-1} = 1 + e^{(\gamma-1)\log(z)}$ is the dimensionless enthalpy function and $f_\infty = f(z_\infty)$, $z_\infty > 0$, its value at infinity. The function $F_\mu$ defined by Eq.~(\ref{Eq:Fmu}) is analytic on the domain $\Omega_c := \{ (x,z)\in\Complex^2 : \re(x) > 0,\re(z) > 0 \}$. In~\cite{eCoS15a} we showed that there exists a unique real-valued differentiable function $z_0: [1,\infty)\to\Real$ (the Michel solution), defined on and outside the event horizon, such that $F_\mu(x,z_0(x)) = f_\infty^2$ for all $x\geq 1$ and $\lim_{x\to\infty} z(x) = z_\infty$. This solution has the property that the partial derivative of $F_\mu$ with respect to $z$,
\begin{equation}
\frac{\partial F_\mu}{\partial z}(x,z) = \frac{2f(z)^2}{z}\nu(z)^2\left[
 1 - \frac{1}{x} - \left( \frac{1}{\nu(z)^2} - 1 \right)\frac{\mu^2}{x^4 z^2} \right],\qquad
\nu := \frac{v_s}{c},
\label{Eq:dFz}
\end{equation}
is different from zero for all $x\geq 1$ except at the location of the critical point $x = x_c$. By continuity, $\partial F_\mu/\partial z$ is also different from zero in an open neighborhood $U\subset \Omega_c$ of the graph $G := \{ (x,z_0(x)) : x\geq 1, x\neq x_c \}$. Therefore, it follows from the implicit function theorem that for an open neighborhood $V\subset U$ of $G$ in $\Omega_c$, $z_0(x)$ admits a unique analytic continuation $z(x)$ whose graph lies in $V$ and such that $F_\mu(x,z(x)) = f_\infty^2$ for all $(x,z(x))\in V$.

It remains to prove that $z(x)$ can be further extended to a neighborhood of $x = \infty$ and to an open neighborhood of the critical point. For the former case, we introduce the new variable $y := 1/x$ and rewrite Eq.~(\ref{Eq:Fmu}) as
$$
\tilde{F}_\mu(y,z) := F_\mu\left( \frac{1}{y},z \right) 
 = f(z)^2\left[ 1 - y + \frac{\mu^2}{z^2} y^4 \right] = f_\infty^2.
$$
The function $\tilde{F}_\mu$ is analytic on the domain $y\in\Complex$, $\re(z) > 0$, and it satisfies $\tilde{F}_\mu(0,z_\infty) = f_\infty^2$ and
$$
\frac{\partial\tilde{F}_\mu}{\partial z}(0,z_\infty)
 = \frac{2f_\infty^2}{z_\infty}\nu(z_\infty)^2\neq 0.
$$
Therefore, it follows from the implicit function theorem that there exists an open neighborhood $\tilde{V}$ of $(0,z_\infty)$ and a unique function $\tilde{z}(y)$ whose graph lies in $\tilde{V}$ such that $\tilde{z}(0) = z_\infty$ and $\tilde{F}_\mu(y,\tilde{z}(y)) = f_\infty^2$ for all $(y,\tilde{z}(y))\in \tilde{V}$. By uniqueness of the analytic continuation, $z(x) = \tilde{z}(1/x)$ for large enough $|x|$, which proves that the analytic extension of $z(x)$ exists for sufficiently large $|x|$. Furthermore,
$$
\lim\limits_{\substack{x\to\infty \\ \re(x) > 0}} z(x) 
 = \lim\limits_{y\to 0} \tilde{z}(y) = z_\infty.
$$

Next, we discuss the analytic continuation of $z_0(x)$ in an open neighborhood of the critical point $x = x_c$. For this, we first note that in a vicinity of the critical point $(x_c,z_c = z_0(x_c))$ the function $F_\mu$ has the Taylor representation
$$
F_\mu(x_c + \xi,z_c + \zeta) = F_\mu(x_c,z_c) 
 + \frac{1}{2}\left[ \frac{\partial^2 F_\mu}{\partial x^2}(x_c,z_c)\xi^2
 + 2\frac{\partial^2 F_\mu}{\partial x\partial z}(x_c,z_c)\xi\zeta
 + \frac{\partial^2 F_\mu}{\partial z^2}(x_c,z_c)\zeta^2 \right] + R_3(\xi,\zeta),
$$
where the error term $R_3(\xi,\zeta)$ is at least cubic in $(\xi,\zeta)$. Let $z_c'\in\Real$ denote one of the two roots of the quadratic polynomial (cf. Eq.~(\ref{Eq:nprimeprime}))
$$
\frac{\partial^2 F_\mu}{\partial x^2}(x_c,z_c)
 + 2\frac{\partial^2 F_\mu}{\partial x\partial z}(x_c,z_c) z_c'
 + \frac{\partial^2 F_\mu}{\partial z^2}(x_c,z_c)(z_c')^2 = 0,
$$
and introduce the function
$$
H_\mu(\xi,\eta) := \left\{ \begin{array}{ll}
\frac{1}{\xi^2}\left[ F_\mu(x_c + \xi, z_c + z_c'\xi \eta ) - F_\mu(x_c,z_c) \right] 
 & \hbox{for $\xi\neq 0$},\\
\frac{1}{2}\left[ \frac{\partial^2 F_\mu}{\partial x^2}(x_c,z_c)
 + 2\frac{\partial^2 F_\mu}{\partial x\partial z}(x_c,z_c)z_c'\eta
 + \frac{\partial^2 F_\mu}{\partial z^2}(x_c,z_c)(z_c')^2\eta^2 \right]
 & \hbox{for $\xi=0$}.
\end{array} \right.
$$
Then, $H_\mu$ is analytic in an open neighborhood of $(\xi,\eta) = (0,1)$ in $\Complex^2$, satisfies $H_\mu(0,1) = 0$ and
$$
\frac{\partial H_\mu}{\partial\eta}(0,1) 
 = \frac{\partial^2 F_\mu}{\partial x\partial z}(x_c,z_c)z_c'
 + \frac{\partial^2 F_\mu}{\partial z^2}(x_c,z_c)(z_c')^2
 = \mp 3\frac{h_c^2}{x_c^3} \frac{\sqrt{1 + 3(\nu_c^2 - W_c)}}{2 \pm \sqrt{1 + 3(\nu_c^2 - W_c)}} \neq 0.
$$
Therefore, using once again the implicit function theorem, it follows the existence of an open neighborhood $Z$ of $(0,1)$ in $\Complex^2$ and a unique analytic function $\eta(\xi)$ whose graph lies inside $Z$ such that $\eta(0) = 1$ and $H_\mu(\xi,\eta(\xi)) = 0$ for all $(\xi,\eta(\xi))\in Z$. By construction $z(x) := z_c + z_c'(x - x_c)\eta(x - x_c)$ is analytic and satisfies $F_\mu(x,z(x)) = F(x_c,z_c) = f_\infty^2$. This demonstrates the existence of the analytic continuation of $z(x)$ in a neighborhood of the critical point.

With these results, it follows directly from Eqs.~(\ref{Eq:DefcalN},\ref{Eq:DefVell},\ref{Eq:nprimeBis}) that the functions ${\cal N}(r)$ and $V_\ell(r)$ have analytic continuations for complex $r$, and that these continuations satisfy Eq.~(\ref{Eq:Limits}).

\bibliographystyle{unsrt}
\bibliography{../References/refs_accretion}

\begin{thebibliography}{10}

\bibitem{EHT}
Event horizon telescope, http://www.eventhorizontelescope.org.

\bibitem{sDetal08}
S.~Doeleman and et~al.
\newblock Event-horizon-scale structure in the supermassive black hole
  candidate at the galactic centre.
\newblock {\em Nature}, 455:78, 2008.

\bibitem{aBtJaLdP14}
A.E. Broderick, T.~Johannsen, A.~Loeb, and D.~Psaltis.
\newblock Testing the no-hair theorem with event horizon telescope observations
  of {Sagittarius A$^*$}.
\newblock {\em Astrophys. J.}, 784:7, 2014.

\bibitem{oZjFlRpM05}
O.~Zanotti, J.A. Font, L.~Rezzolla, and P.J. Montero.
\newblock Dynamics of oscillating relativistic tori around {K}err black holes.
\newblock {\em Mon.Not.Roy.Astron.Soc.}, 356:1371--1382, 2005.

\bibitem{mMetal09}
M.~Megevand, M.~Anderson, J.~Frank, E.W. Hirschmann, L.~Lehner, S.L. Liebling,
  P.M. Motl, and D.~Neilsen.
\newblock Perturbed disks get shocked. binary black hole merger effects on
  accretion disks.
\newblock {\em Phys.Rev.}, D80:024012, 2009.

\bibitem{mAetal10}
M.~Anderson, L.~Lehner, M.~Megevand, and D.~Neilsen.
\newblock Post-merger electromagnetic emissions from disks perturbed by binary
  black holes.
\newblock {\em Phys.Rev.}, D81:044004, 2010.

\bibitem{fM72}
F.C. Michel.
\newblock Accretion of matter by condensed objects.
\newblock {\em Astrophysics and Space Science}, 15:153--160, 1972.

\bibitem{hB52}
H.~Bondi.
\newblock On spherically symmetrical accretion.
\newblock {\em Monthly Notices Roy Astronom. Soc.}, 112:195--204, 1952.

\bibitem{Shapiro-Book}
S.L. Shapiro and S.A. Teukolsky.
\newblock {\em Black Holes, White Dwarfs, and Neutron Stars}.
\newblock John Wiley \& Sons, New York, 1983.

\bibitem{fGfL11}
F.S. Guzm\'an and F.D. Lora-Clavijo.
\newblock Exploring the effects of pressure on the radial accretion of dark
  matter by a {S}chwarzschild supermassive black hole.
\newblock {\em Mon. Not. R. Astron. Soc.}, 415:225--234, 2011.

\bibitem{eCoS15a}
E.~Chaverra and O.~Sarbach.
\newblock Radial accretion flows on static, spherically symmetric black holes.
\newblock 2015.
\newblock arXiv:1501.01641.

\bibitem{vM80}
V.~Moncrief.
\newblock Stability of stationary, spherical accretion onto a {S}chwarzschild
  black hole.
\newblock {\em Astrophys. J.}, 235:1038--1046, 1980.

\bibitem{nB99}
N.~Bilic.
\newblock Relativistic acoustic geometry.
\newblock {\em Class. Quantum Grav.}, 16(12):3953--3964, 1999.

\bibitem{cBsLmV05}
C.~Barcelo, S.~Liberati, and M.~Visser.
\newblock Analogue gravity.
\newblock {\em Living Rev.Rel.}, 8:12, 2005.

\bibitem{pMeM13}
P.~Mach and E.~Malec.
\newblock Stability of relativistic {B}ondi accretion in
  {S}chwarzschild-(anti-)de {S}itter spacetimes.
\newblock {\em Phys.Rev. D}, D88:084055, 2013.

\bibitem{tDnBsD06}
T.K. Das, N.~Bilic, and S.~Dasgupta.
\newblock Black-hole accretion disc as an analogue gravity model.
\newblock {\em JCAP}, 0706:009, 2007.

\bibitem{tD07}
T.K. Das.
\newblock Astrophysical accretion as an analogue gravity phenomena.
\newblock 2007.
\newblock arXiv:0704.3618.

\bibitem{dAsBtD14a}
D.B. Ananda, S.~Bhattacharya, and T.K. Das.
\newblock Acoustic geometry through perturbation of mass accretion rate {I} -
  radial flow in general static spacetime.
\newblock 2014.
\newblock arXiv:1406.4262.

\bibitem{dAsBtD14b}
D.B. Ananda, S.~Bhattacharya, and T.K. Das.
\newblock Acoustic geometry through perturbation of mass accretion rate -
  axisymmetric flow in static spacetimes.
\newblock 2014.
\newblock arXiv:1407.2268.

\bibitem{eBvCjL04}
E.~Berti, V.~Cardoso, and J.~Lemos.
\newblock Quasinormal modes and classical wave propagation in analogue black
  holes.
\newblock {\em Phys. Rev. D}, 70:124006, 2004.

\bibitem{sDlOlC10}
S.R. Dolan, L.A. Oliveira, and Luis L.C.B.~Crispino.
\newblock Quasinormal modes and regge poles of the canonical acoustic hole.
\newblock {\em Phys.Rev.}, D82:084037, 2010.

\bibitem{MQNdeS}
E.~W. Leaver.
\newblock An analytic representation for the quasi-normal modes of {K}err black
  holes.
\newblock {\em Proc. R. Soc Lond. A}, 402:285--29, 1985.

\bibitem{nAcH04}
N.~Andersson and C.J. Howls.
\newblock The asymptotic quasinormal mode spectrum of nonrotating black holes.
\newblock {\em Class. Quantum Grav.}, 21:1623--1642, 2004.

\bibitem{aZhYmZyClL14}
A.~Zimmerman, H.~Yang, Z.~Mark, Y.~Chen, and L.~Lehner.
\newblock Quasinormal modes beyond {K}err.
\newblock {\em Astrophys. Space Sci. Proc.}, 40:217, 2015.

\bibitem{eCoS12}
E.~Chaverra and O.~Sarbach.
\newblock Polytropic spherical accretion flows on {S}chwarzschild black holes.
\newblock {\em AIP Conf.Proc.}, 1473:54--58, 2012.

\bibitem{jKeM13}
J.~Karkowski and E.~Malec.
\newblock Bondi accretion onto cosmological black holes.
\newblock {\em Phys.Rev. D87 (2013)}, D87:044007, 2013.

\bibitem{Heusler-Book}
M.~Heusler.
\newblock {\em Black Hole Uniqueness Theorems}.
\newblock Cambridge University Press, Cambridge, England, 1996.

\bibitem{Wald-Book}
R.M. Wald.
\newblock {\em General Relativity}.
\newblock The University of Chicago Press, Chicago, London, 1984.

\bibitem{hNbS92}
H.-P. Nollert and B.G. Schmidt.
\newblock Quasinormal modes of {S}chwarzschild black holes: Defined and
  calculated via laplace transformation.
\newblock {\em Phys. Rev. D}, 45:2617--2627, 1992.

\bibitem{hN99}
H.-P. Nollert.
\newblock {TOPICAL REVIEW: Quasinormal modes: the characteristic `sound' of
  black holes and neutron stars}.
\newblock {\em Class. Quantum Grav.}, 16:R159--R216, 1999.

\bibitem{kKbS99}
K.D. Kokkotas and B.G. Schmidt.
\newblock Quasi-normal modes of stars and black holes.
\newblock {\em Living Reviews in Relativity}, 2(2), 1999.

\bibitem{eBvCaS09}
E.~Berti, V.~Cardoso, and A.O. Starinets.
\newblock Quasinormal modes of black holes and black branes.
\newblock {\em Class. Quantum Grav.}, 26:163001, 2009.

\bibitem{rN60}
R.G. Newton.
\newblock Analytic properties of radial wave functions.
\newblock {\em J. Math. Phys.}, 1:319--347, 1960.

\bibitem{ReedSimonIII}
M.~Reed and B.~Simon.
\newblock {\em Methods of Modern Mathematical Physics, Vol. III: Scattering
  Theory}.
\newblock Academic Press, San Diego, 1979.

\bibitem{bJpC86}
B.P. Jensen and P.~Candelas.
\newblock The {S}chwarzschild radial functions.
\newblock {\em Phys. Rev. D}, 33:1590--1595, 1986.

\bibitem{eCoS15b}
E.~Chaverra and O.~Sarbach.
\newblock {\em In preparation}, 2015.

\bibitem{Recipes-Book}
W.~H. Press, S.~A. Teukolsky, W.~T. Watterling, and B.~P. Flannery.
\newblock {\em Numerical Recipes in Fortran}.
\newblock Cambridge University Press, Cambridge, 1992.

\bibitem{oSmT12}
O.~Sarbach and M.~Tiglio.
\newblock Continuum and discrete initial-boundary value problems and
  {E}instein's field equations.
\newblock {\em Living Rev. Relativity}, 15, 2012.
\newblock http://www.livingreviews.org/lrr-2012-9.

\bibitem{oSmT01}
O.~Sarbach and M.~Tiglio.
\newblock Gauge invariant perturbations of {S}chwarzschild black holes in
  horizon penetrating coordinates.
\newblock {\em Phys. Rev. D}, 64:084016(1)--(15), 2001.

\bibitem{Chingetal94}
E.S.C. Ching, P.T. Leung, W.M. Suen, and K.~Young.
\newblock Late time tail of wave propagation on curved space-time.
\newblock {\em Phys. Rev. Lett.}, 74:2414--2417, 1995.

\bibitem{vCaMeBhWvZ09}
V.~Cardoso, A.S. Miranda, E.~Berti, H.~Witek, and V.~Zanchin.
\newblock Geodesic stability, {L}yapunov exponents and quasinormal modes.
\newblock {\em Phys. Rev. D}, 79:064016, 2009.

\end{thebibliography}

\end{document}